\documentclass[12pt]{article}
\usepackage{amsmath,amssymb}
\usepackage{amsfonts,mathrsfs}
\usepackage{graphicx}
\usepackage[hypertex]{hyperref}
\usepackage{epsfig}
\numberwithin{equation}{section}

\def\ds{\displaystyle}
\def\ssp{\hspace{0.3mm}}
\def\({\left(}
\def\){\right)}
\renewcommand{\[}{\left[}
\renewcommand{\]}{\right]}
\def\bb#1{\mathbb{#1}}

\def\pare#1{\left\{ #1\right\}}

\def\abs#1{\left| #1\right|}

\def\AdSxS{{AdS${}_5 \times {}$S${}^5$}}

\def\TH{{\rm TH}}

\def\hdl{\, \widehat{dl} \ssp}

\def\ol#1{\overline{#1}}
\renewcommand{\eqref}[1]{$\({\rm \ref{#1}}\)$}

\def\cB{{\mathcal B}}

\def\cN{{\mathcal N}}

\def\cQ{{\mathcal Q}}
\def\cR{{\mathcal R}}

\def\fa{\mathfrak{a}}

\def\fb{\mathfrak{b}}

\def\fX{\mathfrak{X}}

\def\fba{\overline{\mathfrak{a}}}

\def\fbb{\overline{\mathfrak{b}}}

\def\olA{\overline{A}}
\def\olC{\overline{C}}

\def\olF{\overline{F}}

\def\olJ{\overline{J}}

\def\olL{\overline{L}}

\def\olQ{\overline{Q}}
\def\olr{\overline{r}}
\def\olR{\overline{R}}
\def\olU{\overline{U}}

\def\olW{\overline{W}}

\def\olX{\overline{X}}

\def\ov{\over}

\def\hstar{\,\hat{\star}\,}

\def\ts{\tilde{s}}
\newcommand{\alg}[1]{\mathfrak{#1}}
\newcommand{\su}{\alg{su}}

\def\sl(2){\alg{sl}(2)}

\def\be{\begin{equation}}
\def\ee{\end{equation}}
\newcommand{\bea}{\begin{eqnarray}}
\newcommand{\eea}{\end{eqnarray}}

\begin{document}

%%%%%%%%%%%%%%%%%%%%%%%%%
{\ }
\vspace{-10mm}

\begin{flushright}
{\bf January 2011}\\[1mm]
{\small ITP-UU-11-01}\\[1mm]
{\small SPIN-11-01}\\[1mm]
\end{flushright}

\vskip 2cm

\begin{center}
\LARGE
%%%%%%%%%%%%%%%%%%%%%%%%

\mbox{\bf Hybrid NLIE for the Mirror \AdSxS}

%%%%%%%%%%%%%%%%%%%%%%%%

\vskip 2cm
\renewcommand{\thefootnote}{$\alph{footnote}$}

\large
\centerline{\sc Ryo Suzuki\footnotetext{{\tt R.Suzuki@uu.nl}}}

\vskip 1cm

\emph{Institute for Theoretical Physics and Spinoza Institute,\\
Utrecht University, 3508 TD Utrecht, The Netherlands}

\end{center}
%%%%%%%%%%%

\vskip 14mm

\centerline{\bf Abstract}

\vskip 6mm

We revisit the derivation of hybrid nonlinear integral equations of the XXX model starting from the linearization of the T-system related to spinon variables. We obtain two sets of equations, corresponding to two linearly independent solutions of $A_1$ TQ-relation. Recalling that the TQ-relations in the horizontal strips of the $\alg{su}(2|4|2)$-hook is of $A_1$ type, we replace the corresponding Y-functions by a finite number of auxiliary variables.

\vskip 2mm

\vfill
\thispagestyle{empty}
\setcounter{page}{0}
\setcounter{footnote}{0}
\renewcommand{\thefootnote}{\arabic{footnote}}
\newpage

\tableofcontents

\newpage

\section{Introduction and Summary}\label{sec:Intro}

AdS/CFT is a conjecture dictating that the spectrum of \AdSxS\ is dual to the dimension of $D=4,\, \cN=4$ SYM local operators \cite{Maldacena97}. The exact spectrum of both theories have recently been studied intensively based on integrability methods.
One of the central ideas is Thermodynamic Bethe Ansatz (TBA) equations.\footnote{The TBA equations appeared first in \cite{YY69} and generalized to the Heisenberg spin chain in \cite{Gaudin71,Takahashi72}.} The TBA equations are used to compute the exact ground state energy of a finite-volume field theory by interchanging the space and time directions \cite{Zamolodchikov90}.
Such strategy was followed by the study of worldsheet theory with space- and time- coordinates interchanged \cite{AJK05}. This theory is also called mirror \AdSxS\ model, and thoroughly investigated in \cite{AF07}.

The partition function of the mirror \AdSxS\ can be computed by asymptotic Bethe Ansatz \cite{BS05,Beisert05} under the assumption called string hypothesis \cite{Takahashi72,AF09a}.
Variational method is applicable in the thermodynamic limit, and the TBA equations arise as the condition for extremality \cite{BFT09,GKKV09,AF09b}.
The TBA consists of a set of nonlinear integral equations for Y-functions. Each Y-function corresponds to a particular type of boundstates which contribute to the mirror partition function.

In the meantime, Y-system on the $\alg{su}(2|4|2)$-hook has been proposed and solved explicitly in the asymptotic limit \cite{GKV09a}.\footnote{By $\alg{su}(2|4|2)$-hook we always mean the T-hook drawn as Figure \ref{fig:Thook} in Appendix \ref{sec:Y and T}. There are other T-hooks of $\alg{su}(2|4|2)$ type, corresponding to different choices of Kac-Dynkin diagram. \cite{Volin10a,Volin10b}.}
The Y-system is a set of difference equations for Y-functions, and the Y-system can be expressed as T-system, namely another set of difference equations for T-functions \cite{KNS93b,KNS93c}. The T-system is closely related to global symmetry of the given model \cite{GKT10,Volin10b}, and can be derived from simplified TBA equations by the `projection' operator $(\log f) \star s_K^{-1} \equiv \log (f^+ f^-)$ \cite{AF09b, AF09d}.
If proper analyticity data are provided, one can integrate the Y-system to obtain most of the TBA equations \cite{AFSu09,CFT10}, except for the formula for the exact energy.

The original TBA equations capture the exact ground state energy, which is zero in the case of AdS/CFT \cite{FS09}.
A common wisdom to compute the exact spectrum of excited states is to deform the integration contours in the TBA equations, which was observed first in integrable 2D vertex models \cite{KP92} and later in perturbed CFT \cite{DT96, DT97}. Upon contour deformation, the integrals pick up additional terms coming from singularities of Y-functions. As such, the excited-state TBA equations depend on the state and the value of coupling constant under consideration \cite{AFSu09}.
Numerical study also indicates that (the gradient of) the exact energy may exhibit a noticeable change at the critical value of coupling constant \cite{Sergey10conf}.

As applications, the exact dimension of Konishi state has been computed numerically up to $\lambda \approx 664$ \cite{GKV09b} and $\lambda \approx 2046$ \cite{Frolov10}, and it matches string theory results at strong coupling \cite{RT09b,GSSV11,MV11,RT11a}.
Five-loop dimension of Konishi state has been obtained numerically \cite{AFSu10} and analytically \cite{BH10a}, which agrees with the prediction of the generalized L\"uscher formula \cite{Luscher85,JL07,BJ08,BHJL09}, although no field theoretical computation has been done as in the four-loop case \cite{FSSZ07,FSSZ08a,Velizhanin08a}.
Such agreement has been extended to general twist-two states in the $\sl(2)$ sector \cite{LRV09,BH10b}.
At strong coupling, the Y-system can be solved by character formula \cite{KR90}, and the result agrees with the exact energy of semiclassical strings \cite{Gromov09,GKT10}.
See also \cite{GL10,ALT10,ABBN10,dLL10,BLM10} towards the exact spectrum of twisted theories.

Behind these successes of TBA there has always been a huge amount of numerical computation, because infinitely many Y-functions must be determined. Therefore, to simplify the spectral problem we need nonlinear integral equations (NLIE) for finitely many degrees of freedom.
Several types of NLIE are known in other integrable models. For example, in the NLIE of Destri-de Vega (DdV) type, each integral is evaluated on the real axis (or a path around the real axis if there are branch cuts). Such NLIE was derived from Bethe Ansatz equations \cite{KB90,KBP91,DdV92a,DdV92b} or from T-system \cite{KP92,BLZ96a}.
As another example, in the NLIE of Takahashi type, one unknown variable is integrated over a contour on the complex plane \cite{Takahashi00,TSK01}. The Takahashi-type NLIE consists of smaller number of equations than the NLIE of DdV type, but comprehensive understanding of the analyticity data is required for its formulation.
In this paper we are interested in the DdV-type NLIE.

At first sight, it is not even clear if the NLIE for the mirror \AdSxS\ exists at all. A hint for the derivation of NLIE comes from the linearization of T-system by Q-functions \cite{Baxter72}. The linearized equations, or so-called TQ-relations \cite{KLWZ96}, have been studied a lot so far, and are solved by B\"acklund transformation \cite{KSZ07, Hegedus09}, or explicitly by the Wronskians of Q-functions \cite{Tsuboi09a,GKLT10}. Remarkably, in the Wronskian formula all T-functions of $\alg{su}(2|4|2)$-hook are expressed by the Wronskian of eight fundamental Q-functions.
It does not immediately imply, however, that the spectral problem gets simpler. In TBA it is enough to compute each Y-function on the real axis. In the Wronskian formula, each fundamental Q-function has to be evaluated on the whole complex plane. Hence, one more step is needed after having solved the TQ-relations.

Motivated to get something practical, we start investigating a mixture of NLIE and TBA, called hybrid NLIE.
Hybrid NLIE was first formulated in the XXX model at spin $S=k/2$ \cite{Su99}, generalizing earlier successes of \cite{KB90,KBP91}. The idea is to add auxiliary variables to the T-system such that the mixed system of equations is closed within a finite number of variables. The Y-function $Y_k$ is replaced by $(1+Y_k) = (1+\fa) (1+\fba)$, where a pair of variables $\fa, \fba$ represent the degrees of freedom of spinons, and all Y-functions $Y_{j>k}$ disappear from the equations.

Here the term `spinon' is used in the following sense.
According to \cite{Reshetikhin91}, the S-matrix of XXX spin chain at spin $S=k/2$ factorizes into the RSOS R-matrix at level $k$ and the S-matrix of spin $\frac12$ excitations. The former degree of freedom, called RSOS excitation, does not couple to external magnetic field. The latter degree of freedom, called spinon, couples to magnetic field.
It turns out that only the auxiliary fields $\fa, \fba$ are sensitive to magnetic field, while Y-functions are insensitive.
Thus the auxiliary fields are identified as spinons \cite{Su99}.

When $k=1$, there are further intimate relations between the auxiliary fields and `spinons' in the usual sense, that is momentum-carrying excitations over the antiferromagnetic vacuum. For one thing, one can derive the character formula of $\widehat{\alg{sl}(2)}_{k=1}$ as summation over spinons \cite{BPS94,Su98}. For another, as mentioned in \cite{Hegedus03}, the hybrid NLIE of XXX model at $k=1$ reduces to the NLIE of Destri and de Vega \cite{DdV92a,DdV92b}.\footnote{At $k=1$, Y-functions disappear from the hybrid NLIE, and thus the NLIE is no longer hybrid.}
In the DdV approach, the fundamental excitations are usual spinons, and we expect that the auxiliary fields of hybrid NLIE will play the same r\^ole.
More insights have been found in the NLIE's of $O(4)$ sigma model \cite{Hegedus03,GKV08,Caetano10}, sine-Gordon model \cite{DdV92a,DdV92b,DdV94,FMQR96,FRT98} and other integrable lattice models \cite{Klumper93,KWZ93}.

\subsection*{Summary of results}

In this paper, we first revisit the derivation of hybrid NLIE in XXX model \cite{Su99} in a slightly different way.
We start our discussion from new recursion relations for spinon variables, instead of following computations using explicit form of T-functions.
From these lessons we learn how to proceed in the \AdSxS\ case.\footnote{An object called quantum transfer matrix was studied in \cite{Su99}. In AdS/CFT, it is not known how to construct usual or quantum transfer matrices. With a different motivation in mind in contrast to \cite{Su99}, we consider the usual transfer matrix analytically continued to the mirror region as the asymptotic solution of T-system. Despite differences in the physical interpretation, the same mathematical techniques are applicable.}

Next, we repeat the discussions from the TQ-relations which appear in the horizontal strips $(\abs{s}\ge 2)$ of the $\alg{su}(2|4|2)$-hook. We derive a set of NLIE which decomposes the TBA equations for $Y_{M|w}$-strings of the mirror \AdSxS. This set of NLIE can be glued to the other parts of TBA equations together with the exact Bethe equation, and hence our formalism is hybrid. Using analyticity assumptions, the NLIE can be truncated within a finite number of variables.

In the literature, it is commonly recognized that TQ-relations are important to derive DdV-type NLIE (see for instance \cite{BLZ96a}, or appendices of \cite{Klumper92,BH09c}). Our discussions goes almost in parallel, but the derivation is a bit more general. We use the explicit form of T- and Q-functions only for checking analyticity assumptions, and the rest of arguments follows immediately from the symmetry structure of $A_1$ TQ-relations.
Note that, in the integrable models studied so far, there is a gauge choice in which all T-functions on the boundary of $A_1$-type strip are either unity, or some known functions independent of Bethe roots.\footnote{There is another gauge in which the T-function on the boundary is a polynomial of a fixed degree.}
We do not use such conditions in the case of \AdSxS, because it is unclear if there exists such a gauge choice. 

\begin{figure}[tbp]
\begin{center}
\includegraphics[scale=0.6]{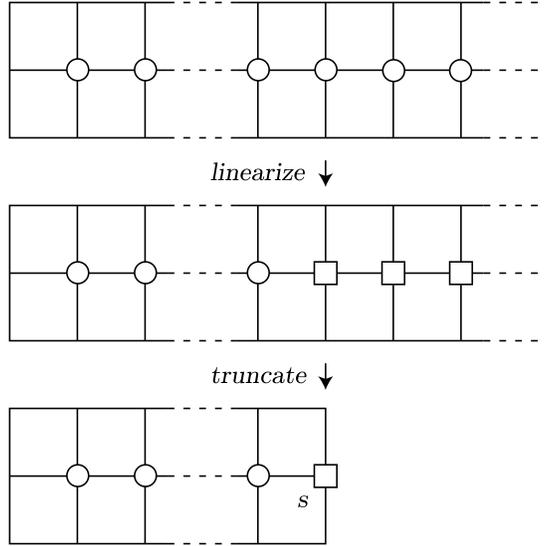}
\caption{The horizontal strip of the $\alg{su}(2|4|2)$-hook. The symbol $\square$ means the TBA equation for the corresponding site is decomposed. Under analyticity assumptions, we can excise the nodes of $\square$.}
\label{fig:Truncate}
\end{center}
\end{figure}

\bigskip
To be specific, let us introduce a set of parameters for the hybrid NLIE as
\begin{equation}
\pare{Y_{1|w} \,, \dots \,, Y_{s-2|w} } \ \bigcup \ 
\pare{ \begin{array}{cccc}
\fa_s^\alpha \\[1mm]
\fba_s^\alpha
\end{array} }, \qquad (s \ge 3),
\label{parameters min s Kon}
\end{equation}
where $\alpha$ is either ${\rm I}$ or ${\rm II}$, and it refers to two linearly independent pairs of Q-functions that solve the TQ-relations.
Our notation is summarized in Appendix \ref{app:notations}.
The pair of variables $(\fa_s^\alpha \,, \fba_s^\alpha)$ satisfy
\begin{equation}
1+Y_{s-1|w} = (1+\fa_s^{\alpha \, [+\gamma]} ) \, (1+\fba_s^{\alpha \, [-\gamma]}),
\label{Y to pp Kon}
\end{equation}
where $\gamma$ is a small parameter for regularization. The parameter $\gamma$ is arbitrary as long as $0<\gamma<1$, and it facilitates numerical computation.
The parametrization \eqref{parameters min s Kon} means that the TBA equation for $Y_{M|w} \ (M \ge s)$ are decomposed into an infinite set of NLIE for auxiliary variables.
Furthermore, under the assumptions on analyticity \eqref{analytic T1s-1}, \eqref{analytic cQL} and \eqref{analytic 1+bs}, we can truncate the NLIE for auxiliary variables at finite $s$. This structure is summarized in Figure \ref{fig:Truncate}

\bigskip
Below we summarize the minimal set of NLIE's in the case of Konishi state in the $\sl(2)$ at weak coupling.
We use the mirror TBA equations to determine $Y_Q \,, Y_{M|vw} \,, Y_\pm$, and also the exact Bethe equations to determine Bethe roots \cite{GKKV09,AFSu09}.
The simplified TBA for $Y_{1|w}$ is given in \cite{AFSu09}, which we rewrite as
\begin{equation}
\log Y_{1|w} = \log (1+\fa_3^{\alpha \, [+\gamma]} ) \, (1+\fba_3^{\alpha \, [-\gamma]}) \star s_K
+ \log \frac{1-\frac{1}{Y_-} }{1-\frac{1}{Y_+} } \hstar s_K \,.
\label{TBA Y1w Kon}
\end{equation}
The pair of parameters $(\fa_3^\alpha \,, \fba_3^\alpha)$ are determined by
\begin{align}
\log \fa_3^\alpha &= \log (1+\fa_3^\alpha) \star K_f
- \log (1+\fba_3^\alpha) \star K_f^{[+2-2\gamma]}
+ \log (1+Y_{1|w}^{[-\gamma]} ) \star s_K + J_3^\alpha \,,
\label{p NLIEa Kon} \\[2mm]
\log \fba_3^\alpha &= \log (1+\fba_3^\alpha) \star K_f 
- \log (1+\fa_3^\alpha) \star K_f^{[-2+2\gamma]}
+ \log (1+Y_{1|w}^{[+\gamma]} ) \star s_K + \olJ_3^\alpha \,,
\label{p NLIEb Kon}
\end{align}
where $J_3^{\rm I} = \olJ_3^{\rm I} = 0$ and
\begin{align}
J_3^{\rm II} &= - \log S_f (v) - \log S_f \Big( v + \frac{2i \, (1-\gamma)}{g} \Big) - \pi i,
\notag \\[2mm]
\olJ_3^{\rm II} &= + \log S_f (v) + \log S_f \Big( v - \frac{2i \, (1-\gamma)}{g} \Big) + \pi i,
\label{source Kon}
\end{align}
with possibly a multiple of $2\pi i$ in addition. The source terms \eqref{source Kon} come from the branch cut discontinuity of $\log (1+\fa_3^{\rm II}) \,, \log (1+\fba_3^{\rm II})$ at the origin.
These equations can be checked by using the asymptotic solution shown in Appendix \ref{app:asymp sol}.

\bigskip
This paper is organized as follows.
In Section \ref{sec:NLIE XXX}, we review the derivation of hybrid NLIE in the XXX model, with a slight modification from \cite{Su99}.
In Section \ref{sec:exact 1s}, we derive hybrid NLIE from TQ-relations in the horizontal strips of $\alg{su}(2|4|2)$-hook, which is of $A_1$ type.
Section \ref{sec:discussion} is for discussions.
In appendices, we summarize the notation, the relation between TBA, Y- and T-system of \AdSxS, and the asymptotic solutions of recursion relations discussed in the main text.

\section{Hybrid NLIE in XXX model}\label{sec:NLIE XXX}

As a warm-up, we revisit the derivation of hybrid NLIE in the XXX model at spin $k/2$ \cite{Su99}.

\subsection{Quantum transfer matrix in spin $k/2$ XXX model.}

The quantum transfer matrix (QTM) is a method to compute the free-energy of spin $k/2$ XXX model at finite temperature without string hypothesis \cite{Suzuki85,Klumper92}.
In this method, we first map the XXX model of size $L$ to two-dimensional vertex model of size $L \times R$, where $R$ is called Trotter number \cite{Suzuki85,Klumper92}. The time (or temperature) direction is replaced by a discrete spatial direction. Then we consider states which look like $M$-particle state over ferromagnetic vacuum, and take the Trotter limit $R \to \infty$.
In this limit, the largest eigenvalue of QTM $T_k$ is equal to the free-energy at finite temperature, under appropriate identification of parameters. For instance, the number $M$ also goes to infinity, with $M/R=k/2$ fixed.

Just like ordinary transfer matrix, QTM depends on the `mirror' Bethe roots $x_1 \,, x_2 \,, \dots x_M$\,.
Since we do not postulate string hypothesis, we do not know their exact position.
Instead, we compute the largest eigenvalue of QTM in the Trotter limit by solving a set of NLIE.
We relegate the precise definition of QTM to \cite{Su99,EFGKKtext} and references therein, as we are interested in the algebraic structure, rather than the physical origin, of QTM.

Let us use the notation $f^{[+Q]} = f (v+iQ),\ f^\pm = f^{[\pm 1]}$.
The elementary QTM $T_1 (v)$ reads \cite{Su99}
\begin{gather}
T_1 (v) = e^{\beta H} \, \phi_+^{[-k+1]} \, \phi_-^{[-k-1]} \, \frac{Q^{[+2]} }{Q }
+ e^{-\beta H} \, \phi_+^{[k+1]} \, \phi_-^{[k-1]} \, \frac{Q^{[-2]} }{Q} \,,
\label{QT1 XXX} \\[1mm]
\phi_\pm (v) = (v \pm iu)^{R/2} ,
\qquad
Q (v) = \prod_{i=1}^M (v-x_i).
\label{def:Q XXX}
\end{gather}
The QTM for higher representations satisfy the T-system equations (also called functional relations among the fusion hierarchy) \cite{KP92, KNS93b, KNS93c},
\begin{align}
T_j^+ \, T_j^- = T_{j-1} \, T_{j+1} + f_j \,, \qquad
f_j = \prod_{m=1}^j \prod_{\sigma=\pm} \phi_\sigma^{[\sigma (j-k-2m+1)]}
\phi_\sigma^{[\sigma (-j+k+2m+1)]} \,,
\label{T system XXX}
\end{align}
with $j \ge 1$ and $T_0 = 1$. The temperature is related to the parameter $u$, which goes to zero in the Trotter limit.
Explicit form of $T_j (v)$ reads \cite{Su99}
\begin{align}
T_j (v) &= \sum_{\ell=1}^{j+1} \lambda_j^{(k)} (v,\ell)
= \sum_{\ell=1}^{j+1} e^{\beta H (j+2-2\ell)} \, \psi_j^{(k)} (v,\ell) \,
\frac{Q^{[j+1]} \, Q^{[-j-1]} }{Q^{[2 \ell -j-1]} \, Q^{[2 \ell -j-3]} } \,,
\label{QTj XXX} \\[1mm]
\psi_j^{(k)} (v,\ell) &= \prod_{m=1}^{j-\ell+1} \phi_-^{[j-k-2m]} \, \phi_+^{[j-k+2-2m]} \times
\prod_{m=1}^{\ell-1} \phi_-^{[-j+k-2+2m]} \, \phi_+^{[-j+k+2m]} \,,
\notag
\end{align}
The T-system \eqref{T system XXX} is invariant under the gauge transformation
\begin{equation}
T_j \ \to \ g_1^{[+j]} \, g_2^{[-j]} \, T_j \,,\qquad
f_j \ \to \ g_1^{[j-1]} \, g_1^{[j+1]} \, g_2^{[-j+1]} \, g_2^{[-j-1]} \, f_j \,.
\label{Tj gauge trs}
\end{equation}

\subsection{Spinon variables}\label{sec:spinon XXX}

We define auxiliary variables by the top and the bottom component of the sum in \eqref{QTj XXX},
\begin{alignat}{5}
A_j (v) &= \lambda_{j}^{(k)} (v,j+1) & &= e^{-j \beta H} \, \psi_j^{(k)} (v,j+1) \, \frac{Q^{[-j-1]} }{Q^{[+j-1]} } \,,
\notag \\[1mm]
\olA_j (v) &= \lambda_j^{(k)} (v,1) & &= e^{+j \beta H} \, \psi_j^{(k)}(v,1) \, \frac{Q^{[+j+1]} }{Q^{[-j+1]} } \,.
\label{def:A's XXX}
\end{alignat}
By $X$ and $\ol{X}$ we denote independent degrees of freedom. We will call them conjugate, as they are sometimes (but not always) complex conjugate with each other.

It is straightforward to check the following pair of recursion relations
\begin{equation}
T_j^+ - A_j^+ = \frac{\olA_j^+}{\olA_{j-1}} \, T_{j-1} \,,
\qquad
T_j^- - \olA_j^- = \frac{A_j^-}{A_{j-1}} \, T_{j-1} \,,
\qquad (j \ge 1),
\label{TA rec XXX}
\end{equation}
with $A_0 = \olA_0 = T_0 = 1$. To maintain the full gauge symmetry of the T-system, we require that $A_j \,, \olA_j$ behave in the same way as $T_j$ under the gauge transformation \eqref{Tj gauge trs}.
The ratio of two gauge-covariant variables are gauge-invariant,
\begin{equation}
\frac{T_j^+ }{A_j^+ } = 1 + \fb_j \,,
\qquad
\frac{T_j^- }{\olA_j^- } = 1 + \fbb_j \,.
\label{def:fbb}
\end{equation}
Following the discussion of \cite{Su99}, we call $A_j \,, \olA_j$ or $\fb_j \,, \fbb_j$ spinon variables, and \eqref{TA rec XXX} covariant recursions for spinons. In the literature, the right hand side of \eqref{TA rec XXX} was recognized only as $({\rm something}) \times T_{j-1}$\,, with `something' determined case by case. The new recursions \eqref{TA rec XXX} are the basis of the following discussions in this section.

\bigskip
The pair of recursions \eqref{TA rec XXX} with the definition of auxiliary variables \eqref{def:A's XXX} appear in various integrable models.
For example, the QTM \eqref{QT1 XXX}, \eqref{QTj XXX} look similar to the ordinary transfer matrix of the XXX model and the $O(4)$ model \cite{KR87,Hegedus03}. They differ only by the form of the function $\psi_j^{(k)} (v, \ell)$. Not surprisingly, the pair of covariant recursions \eqref{TA rec XXX} are satisfied with the definition of \eqref{def:A's XXX}.

In fact, the recursions \eqref{TA rec XXX} follow from $A_1$ TQ-relations, as discussed later in Section \ref{sec:exact 1s}.\footnote{The author thanks Zoltan Bajnok for these remarks and suggestion for this identification.}
Let us identify
\begin{equation}
A_j^\alpha = \frac{\olQ_{j-1}^{\alpha \,-} }{Q_{j-1}^{\alpha \,-} } \, L_j \,,
\qquad
\olA_j^\alpha = \frac{Q_{j-1}^{\alpha \,+} }{\olQ_{j-1}^{\alpha \,+} } \, \olL_j \,,
\qquad (\alpha = {\rm I}, {\rm II}),
\label{relation AQL XXX}
\end{equation}
where the index $\alpha$ refers to two linearly independent solutions of $A_1$ TQ-relations.
In the (twisted) XXX model, the two sets of Q-functions are given by a polynomial of Bethe roots and holes as \eqref{def:Q XXX} \cite{PS98,Pronko99,BLMS10}.\footnote{The term `hole' refers to the excitations from the highest energy state of ferromagnetic nature, and not to the excitations from the antiferromagnetic state.}
The pair $(A_j \,, \olA_j)$ shown in \eqref{def:A's XXX} should correspond to the case of Bethe roots.

As shown in Appendix \ref{app:asymp sol}, one can solve the recursions \eqref{TA rec XXX} by using the asymptotic transfer matrix $T_{a,1}$ in the $\alg{sl}(2)$ sector, with the definition of $A$'s similar to \eqref{def:A's XXX}.
This also means that the asymptotic transfer matrix $T_{1,s}$ in the $\alg{su}(2)$ sector satisfies the recursions in the same way, because the transfer matrices in the $\sl(2)$- and $\alg{su}(2)$-sectors are related by the flip $T_{a,s} \leftrightarrow T_{s,a}$ with the interchange $\pm \leftrightarrow \mp$.

\subsection{Decomposing the Y-system of XXX model}\label{sec:decomp XXX}

Let us discuss the consequences of the covariant recursions for spinons \eqref{TA rec XXX}.

\smallskip
First, using \eqref{TA rec XXX}, we express $T_{j+1}$ by $T_{j-1}$ in two ways. The first expression is given through $T_{j+1} \to T_j^- \to T_{j-1}$\,, while the second expression is through $T_{j+1} \to T_j^+ \to T_{j-1}$\,. Since both results are equivalent, we obtain the compatibility condition
\begin{equation}
\Delta (A_j) = \Delta (\olA_j) \equiv 1 + \fX_j \,,
\qquad
\Delta (X_j) \equiv \frac{X_j^- \, X_j^+}{X_{j-1} \, X_{j+1} } \,.
\label{Laplace Aj}
\end{equation}
The operator $\Delta$ is called discrete Laplacian.
When this condition is satisfied, we obtain
\begin{equation}
T_j^- \, T_j^+ - (1+\fX_j) \, T_{j-1} \, T_{j+1} = A_j^+ \, \olA_j^- \,,
\label{Tjpm ev}
\end{equation}
which is equivalent to the T-system \eqref{TA rec XXX} if
\begin{equation}
\fX_j = 0, \qquad f_j = A_j^+ \, \olA_j^-\,.
\end{equation}
We set $\fX_j$ to zero below, which is indeed true in all examples mentioned so far. Only then, the recursion pair \eqref{TA rec XXX} can be recognized as the linearization of the T-system.
In the gauge-invariant language, the equation \eqref{Tjpm ev} translates into
\begin{equation}
(1+\fb_j) \, (1+\fbb_j) = 1+Y_j \,,
\qquad
1+Y_j \equiv \frac{T_j^- \, T_j^+}{f_j} \,.
\label{1+B to Y exact}
\end{equation}

Second, we rewrite the recursion in terms of $\fb$'s as
\begin{equation}
\fb_j = \frac{\olA_j^+}{A_j^+} \, \frac{T_{j-1}}{\olA_{j-1}} = \frac{1+\fb_j }{1+\fbb_j^{[+2]} } \, (1+\fbb_{j-1}^+),
\qquad
\fbb_j = \frac{A_j^-}{\olA_j^-} \, \frac{T_{j-1}}{A_{j-1}} = \frac{1+\fbb_j}{1+\fb_j^{[-2]} } \, (1+\fb_{j-1}^-).
\label{B to AA}
\end{equation}
By taking the ratio $({\rm LHS})_{j+1}/({\rm LHS})_j^\mp$ we find the following recursion,
\begin{align}
1+\frac{1}{\fb_{j+1}^-} = \frac{1}{1+\frac{1}{Y_j} } \( 1+\frac{1}{\fb_j } \),
\qquad
1+\frac{1}{\fbb_{j+1}^+} = \frac{1}{1+\frac{1}{Y_j} } \( 1+\frac{1}{\fbb_j } \).
\label{Bj recursion}
\end{align}
Using \eqref{1+B to Y exact} and \eqref{Bj recursion}, we can derive alternative expressions of gauge-invariant recursions,
\begin{align}
\fb_j \, (1+\fbb_{j+1}^+) = \fbb_j \, (1+\fb_{j+1}^-) = Y_j \,.
\label{Bj recursion2}
\end{align}

\bigskip
Let us count the number of gauge-invariant variables and the number of equations. The list of gauge-invariant variables is given by
\begin{equation}
\pare{Y_1 \,, Y_2 \,, \dots \,, Y_{j-1} } \ \bigcup \ 
\pare{ \begin{array}{cccc}
\fb_j \\[1mm]
\fbb_j
\end{array} }.
\label{parameters XXX1}
\end{equation}
We define Y-functions by
\begin{alignat}{7}
1+Y_m &= \frac{T_m^- \, T_m^+}{f_m} &\qquad &(m=1,2, \dots, j-1),
\notag \\[1mm]
1+Y_m &= (1+\fb_m) \, (1+\fbb_m) &\qquad &(m=j),
\label{def:Yj XXX2}
\end{alignat}
where $j \ge 1$ is an arbitrary positive integer. 
Then, the T-system at site $m$ is equivalent to the Y-system at site $m$,
\begin{align}
Y_m^- \, Y_m^+ &= (1+Y_{m-1}) \, (1+Y_{m+1}) \qquad (m=1,2, \dots, j-2),
\notag \\[1mm]
Y_{j-1}^- \, Y_{j-1}^+ &= (1+Y_{j-2}) \, (1+\fb_j) \, (1+\fbb_j),
\label{Ysystem XXX2}
\end{align}
where we used $Y_0 = 0$ and $f_m^- \, f_m^+ = f_{m-1} \, f_{m+1}$\,.
There are $j-1$ Y-functions and a pair of auxiliary variables, while there are only $j-1$ equations in \eqref{Ysystem XXX2}. Therefore two equations are missing.

One equation comes from the product of \eqref{B to AA} as
\begin{equation}
\fb_j^- \, \fbb_j^+ = (1+\fb_{j-1}) \, (1+\fbb_{j-1}) = 1+Y_{j-1} \,.
\label{RecRec XXX}
\end{equation}
From algebraic point of view, there is no other way to relate $(\fb_j \,, \fbb_j)$ with $Y_{j-1}$\,.
A closed set of equations can be obtained with the help of analyticity assumptions, which will be reviewed in Section \ref{sec:ANZC XXX}.

\bigskip
One may try to enclose the equations by extending the parameter list, like
\begin{equation}
\pare{Y_1 \,, Y_2 \,, \dots \,, Y_{j-1} } \ \bigcup \ 
\pare{ \begin{array}{cccc}
\fb_j & \fb_{j+1} \\[1mm]
\fbb_j & \fbb_{j+1} \\[1mm]
\end{array} }.
\label{parameters XXX2}
\end{equation}
Now two more variables $(\fb_{j+1} \,, \fbb_{j+1})$ are introduced compared to \eqref{parameters XXX1}. They are constrained by the recursion \eqref{Bj recursion}.
Actually it is possible to find more than two equations: the equation \eqref{RecRec XXX} at $(j,j+1)$, the discrete Laplace equation for $(1+\fb_j)$, and the Y-system at site $j$. However, these extra equations never bring new constraints, because all of them follow from the algebraic relations we have already used.
For example, the equation
\begin{equation}
\fb_{j+1}^- \, \fbb_{j+1}^+ = (1+\fb_j) \, (1+\fbb_j) = 1+Y_j \,,
\label{RecRec XXX j+1}
\end{equation}
is a corollary of \eqref{1+B to Y exact} and \eqref{Bj recursion}. The discrete Laplacian $\Delta (1+\fb_j)=1+1/Y_j^+$ is a corollary of two recursions \eqref{B to AA}, \eqref{Bj recursion2}. The Y-system at site $j$ follows from \eqref{Bj recursion2}, \eqref{RecRec XXX} as
\begin{equation}
Y_j^- \, Y_j^+ = \fb_j^- \, (1+\fbb_{j+1}) \, \fbb_j^+ \, (1+\fb_{j+1}) = (1+Y_{j-1}) \, (1+Y_{j+1}).
\end{equation}
Therefore, after having pondered over the parameter list \eqref{parameters XXX2}, we again find that one variable is yet undetermined.

Similarly, we cannot get the one missing equation even if we extend the parameter list to
\begin{equation}
\pare{Y_1 \,, Y_2 \,, \dots \,, Y_{j-1} } \ \bigcup \ 
\pare{ \begin{array}{cccc}
\fb_j & \fb_{j+1} & \dots & \fb_{j+\ell} \\[1mm]
\fbb_j & \fbb_{j+1} & \dots & \fbb_{j+\ell} \\[1mm]
\end{array} },
\label{parameters XXX3}
\end{equation}
as long as $\ell$ is finite. When we take the limit $\ell \to \infty$, we will just find that the Y-system (or TBA equations) for $Y_{m \ge j}$ are decomposed into the NLIE for spinon variables.

\subsection{Analyticity conditions in the XXX model}\label{sec:ANZC XXX}

We derive two more equations from analyticity assumptions, following \cite{Su99}.

Let us evaluate the first equalities of \eqref{B to AA} by using the explicit form of $A$'s \eqref{def:A's XXX},
\begin{alignat}{5}
\fb_j &= e^{+(j+1) \beta H} \, \frac{\psi_j^{(k)}(v+i,1)}{\psi_j^{(k)}(v+i,j+1) \, \psi_{j-1}^{(k)}(v,1)} \,
\frac{Q^{[+j+2]} }{Q^{[-j]} } \, T_{j-1}
& &\equiv W_j \, \frac{Q^{[+j+2]} }{Q^{[-j]} } \, T_{j-1} \,,
\label{fbj ev1} \\[1mm]
\fbb_j &= e^{-(j+1) \beta H} \, \frac{\psi_j^{(k)}(v-i,j+1)}{\psi_j^{(k)}(v-i,1) \, \psi_{j-1}^{(k)}(v,j)} \,
\frac{Q^{[-j-2]} }{Q^{[+j]} } \, T_{j-1}
& &\equiv \olW_j \, \frac{Q^{[-j-2]} }{Q^{[+j]} } \, T_{j-1} \,.
\label{fbj ev2}
\end{alignat}
We want to take the Fourier transform of the logarithmic derivative of these equations, denoted as
\begin{equation}
\hdl f \equiv \int_{-\infty}^\infty \, dv \, e^{iqv} \, \frac{\partial}{\partial v} \log f(v).
\label{def:hdl}
\end{equation}
The Fourier transform is well-defined only if there are no poles nor branch cuts over the path of integration. As for the left hand side of \eqref{fbj ev1}, \eqref{fbj ev2}, we introduce the regularization
\begin{equation}
\fa_j (v) = \fb_j (v-i \gamma), \qquad
\fba_j (v) = \fbb_j (v+i \gamma),
\label{def:fa}
\end{equation}
with $\gamma>0$ a small parameter, and take the Fourier transform on the real axis of $(\fa_j (v) \,, \fba_j (v))$. As for the right hand side, we assume the analyticity of each factor in \eqref{fbj ev1} for $-\gamma \le {\rm Im} \, v \le 0$, and similarly in \eqref{fbj ev2} for $0 \le {\rm Im} \, v \le +\gamma$.\footnote{These assumptions are part of the analyticity conditions discussed later in \eqref{analytic Tj}, \eqref{analytic QQ}.}
Then these equations become
\begin{align}
\hdl \fa_j &= e^{- \gamma q} \pare{ \hdl W_j + \hdl Q^{[+j+2]} - \hdl Q^{[-j]} + \hdl T_{j-1} },
\notag \\[1mm]
\hdl \fba_j &= e^{+ \gamma q} \pare{ \hdl \olW_j + \hdl Q^{[-j-2]} - \hdl Q^{[+j]} + \hdl T_{j-1} }.
\label{fbj ev Fourier}
\end{align}
We will see later that $(\hdl Q^{[+j]} \,, \hdl Q^{[-j]})$ are related to $(\hdl (1+\fb_j) \, \hat s_K (q), \hdl (1+\fbb_j) \, \hat s_K (q))$, where
\begin{equation}
\hat s_K (q) = \int_{-\infty}^\infty \, dv \, e^{iqv} \, s_K (v) = \frac{1}{2 \cosh q} \,.
\label{def:hat sK}
\end{equation}
Here we rescaled the kernel $s_K$ defined in \eqref{def:sK} by a factor of $g$, because $f^\pm = f(v \pm i)$ in the XXX model.

\bigskip
The terms $\hdl Q^{[+j+2]}$ and $\hdl Q^{[-j-2]}$ in \eqref{fbj ev Fourier} are dangerous in the following sense. Under proper conditions, one may analytically continue them as
\begin{equation}
\hdl Q^{[+j+2]} = e^{+2q} \hdl Q^{[+j]} \,, \qquad
\hdl Q^{[-j-2]} = e^{-2q} \hdl Q^{[-j]} \,.
\end{equation}
Then, the corresponding terms in \eqref{fbj ev Fourier} behave, in the limit ${\rm Re} \; q \to \pm \infty$, as
\begin{alignat}{7}
e^{+2q} \hdl Q^{[+j]} &\sim \hdl (1+\fb_j) \, \hat s_K (q) \, e^{+2q} &\; &\to \; \frac12 \, \hdl (1+\fb_j) \, e^{+2q-\abs{q}} \,,
\notag \\[2mm]
e^{-2q} \hdl Q^{[-j]} &\sim \hdl (1+\fbb_j) \, \hat s_K (q) \, e^{-2q} &\; &\to \; \frac12 \, \hdl (1+\fbb_j) \, e^{-2q-\abs{q}} \,,
\label{danger Qj}
\end{alignat}
which are exponentially growing. If one applies the inverse Fourier transform, one finds that the functions $(\log (1+\fb_j), \log (1+\fbb_j))$ are convoluted with the kernel which diverges exponentially as ${\rm Re} \, v \to \pm \infty$.
As discussed below, we use analyticity conditions such that both $\hdl Q^{[+j]}$ and $\hdl Q^{[-j]}$ vanish in the region where the kernels are exponentially growing.

\subsubsection*{ANZC conditions}

We fix a gauge for the symmetry of T-system \eqref{Tj gauge trs} by the explicit form of $T_j$ in \eqref{QTj XXX}, and assume that in this gauge T-functions are analytic, nonzero and constant at infinity (ANZC), namely
\begin{equation}
T_j (v) \quad \text{is  ANZC  for} \quad -1 \le \,{\rm Im} \, v \le 1,
\qquad 1 \le j \le k.
\label{analytic Tj}
\end{equation}
The upper bound for $j$ comes from an empirical observation that $T_j (v)$ may have zeroes around the lines ${\rm Im} \, v = \pm (k+2-j)$ \cite{Su99}.
Let $D$ be a strip $\abs{{\rm Im} \, v} \le 1$, and consider an integral running along the boundary of $D$. By virtue of the ANZC for $\[ \log T_j \]'$, we find
\begin{align}
0 &= \int_{\partial D} \, dv \, e^{iqv} \frac{\partial}{\partial v} \log T_j (v)
= e^{+q} \hdl T_j (v-i) - e^{-q} \hdl T_j (v+i) \,,
\notag \\[1mm]
&= e^{+q} \hdl \Big( (1+\fbb_j (v) ) \, \olA_j (v-i) \Big) - e^{-q} \hdl \Big( (1+\fb_j (v)) \, A_j (v+i) \Big).
\end{align}
We assume that the Fourier transform of the logarithmic derivative of each factor is unambiguous, namely $1+\fb_j (v), 1+\fbb_j (v), A_j (v+i)$ and $\olA_j (v-i)$ do not have poles nor branch cuts on the real axis of $v$. Then we obtain\footnote{Here $A_j (v+i)$ is equivalent to $A_j^+ = A_j (v+i-i0)$ thanks to the analyticity assumption; similarly for $\olA_j^-$.}
\begin{equation}
0 = e^{+q} \hdl (1+\fbb_j)
- e^{-q} \hdl (1+\fb_j)
+ e^{+q} \hdl \olA_j^-
- e^{-q} \hdl A_j^+ .
\label{analyticity Tj 2}
\end{equation}
The last two terms can be explicitly evaluated by using \eqref{def:A's XXX}, as
\begin{align}
0 &= e^{+q} \hdl (1+\fbb_j) - e^{-q} \hdl (1+\fb_j) + 2 \ssp \cosh q \( \hdl Q^{[+j]} - \hdl Q^{[-j]} \) + \omega_j (q) \,,
\label{analyticity Tj 3} \\[1mm]
\omega_j (q) &\equiv e^{+q} \hdl \psi_j^{(k)}(v-i,1) - e^{-q} \hdl \psi_j^{(k)} (v+i,j+1) .
\label{def:omega}
\end{align}

Next we assume that, for $j \ge k$,
\begin{equation}
Q^{[+j]} (v) \quad \text{is ANZC for} \ \ {\rm Im} \, v \ge 0
\qquad
Q^{[-j]} (v) \quad \text{is ANZC for} \ \ {\rm Im} \, v \le 0.
\label{analytic QQ}
\end{equation}
By closing the contour of Fourier integral over the upper or lower half plane, we find
\begin{equation}
\hdl Q^{[+j]} (q) = 0 \quad {\rm for} \ \ {\rm Re} \; q > 0,
\qquad
\hdl Q^{[-j]} (q) = 0 \quad {\rm for} \ \ {\rm Re} \; q < 0.
\end{equation}
The assumption $j \ge k$ comes from another empirical fact. It is expected that the zeroes of Q-functions lie around ${\rm Im} \, v = k-1, k-3, \dots -k+1$, forming $k$-strings. If so, the Q-functions are ANZC only outside the strip $-k+1 \le {\rm Im} \, v \le k-1$ \cite{Su99}.
Now by applying the analyticity \eqref{analytic QQ} to \eqref{analyticity Tj 3}, we obtain for $j \ge k$,
\begin{alignat}{5}
\hdl Q^{[+j]} (q) &= 
\begin{cases}
0 &\qquad {\rm Re} \; q > 0, \\
\( + e^{-q} \hdl (1+\fb_j) - e^{+q} \hdl (1+\fbb_j) - \hdl \omega_j \) \hat s_K (q) &\qquad {\rm Re} \; q < 0,
\end{cases}
\notag \\[2mm]
\hdl \olQ^{[-j]} (q) &=
\begin{cases}
\( - e^{-q} \hdl (1+\fb_j) + e^{+q} \hdl (1+\fbb_j) + \hdl \omega_j \) \hat s_K (q) &\qquad {\rm Re} \; q > 0, \\
0 &\qquad {\rm Re} \; q < 0.
\end{cases}
\label{QQ to Lambda}
\end{alignat}
When $q=0$, the Fourier transform \eqref{def:hdl} is equal to the difference
\begin{equation*}
\hdl f (q=0) = \log f(+\infty) - \log f(-\infty),
\end{equation*}
which is not so important in computing the inverse Fourier transform.

For later use, we apply the regularization \eqref{def:fa} to $(1+\fb_j) \,, (1+\fbb_j)$. Provided that
\begin{alignat}{7}
1+\fb_j (v) \ \ &\text{is ANZC for} &\ \ -\gamma \le \, &{\rm Im} \, v \le 0,
\notag \\[1mm]
1+\fbb_j (v) \ \ &\text{is ANZC for} &\ \ 0 \le \, &{\rm Im} \, v \le +\gamma,
\end{alignat}
we obtain
\begin{equation}
\hdl (1+\fb_j) = e^{+ \gamma q} \hdl (1+\fa_j), \qquad
\hdl (1+\fbb_j) = e^{- \gamma q} \hdl (1+\fba_j).
\label{def:fa Fourier}
\end{equation}

\subsubsection*{Derivation of NLIE}

We shall set $j=k$ to comply with the ANZC conditions of \eqref{analytic Tj} and \eqref{analytic QQ}. Let us rewrite the T-function in the equations \eqref{fbj ev Fourier} by Y-functions using \eqref{analytic Tj} and \eqref{def:Yj XXX2}, as
\begin{alignat}{7}
e^{+ \gamma q} \hdl \fa_k &= \hdl W_k + e^{+2q} \hdl Q^{[+k]} - \hdl Q^{[-k]} + \pare{ \hdl f_{k-1} + \hdl (1+Y_{k-1}) } \hat s_K (q),
\notag \\[1mm]
e^{-\gamma q} \hdl \fba_k &= \hdl \olW_k + e^{-2q} \hdl Q^{[-k]} - \hdl Q^{[+k]} + \pare{ \hdl f_{k-1} + \hdl (1+Y_{k-1}) } \hat s_K (q),
\label{fbk ev Fourier2}
\end{alignat}
with $f_{k-1} = A_{k-1}^+ \, \olA_{k-1}^- = \psi_{k-1}^{(k)} (v+i,k) \, \psi_{k-1}^{(k)} (v-i,1)$. We then substitute the results \eqref{QQ to Lambda}, \eqref{def:fa Fourier} into these equations, and find
\begin{align}
\hdl \fa_k &= \frac{e^{-\abs{q}} }{2 \ssp \cosh q} \, \hdl (1+\fa_k)
- \frac{e^{2q (1-\gamma) - \abs{q}} }{2 \ssp \cosh q} \, \hdl (1+\fba_k)
+ \frac{e^{-\gamma q} }{2 \ssp \cosh q} \, \hdl (1+Y_{k-1})
\notag \\[2mm]
&\qquad + e^{-\gamma q} \hdl W_k + \frac{e^{-\gamma q} }{2 \ssp \cosh q} \, \hdl f_{k-1}
+ \frac{e^{q (1-\gamma) - \abs{q}} }{2 \ssp \cosh q} \, \hdl \omega_k \,,
\label{XXX nlie a} \\[3mm]
\hdl \fba_k &= \frac{e^{-\abs{q}} }{2 \ssp \cosh q} \, \hdl (1+\fba_k)
- \frac{e^{-2q (1-\gamma) - \abs{q}} }{2 \ssp \cosh q} \, \hdl (1+\fa_k)
+ \frac{e^{-\gamma q} }{2 \ssp \cosh q} \, \hdl (1+Y_{k-1})
\notag \\[2mm]
&\qquad + e^{+\gamma q} \hdl \olW_k + \frac{e^{+\gamma q} }{2 \ssp \cosh q} \, \hdl f_{k-1}
+ \frac{e^{-q (1-\gamma) - \abs{q}} }{2 \ssp \cosh q} \, \hdl \omega_k \,.
\label{XXX nlie b}
\end{align}
The second lines of \eqref{XXX nlie a}, \eqref{XXX nlie b} are known functions in the XXX model, so we can regard all of them as part of the source terms.\footnote{These extra terms will be studied carefully in Section \ref{sec:exact 1s}.}
By applying the inverse Fourier transform, we obtain the missing equations for hybrid NLIE. The results can be summarized as,
\begin{align}
\log \fa_k &= \log (1+\fa_k) \star K_f - \log (1+\fba_k) \star K_f^{[+2-2\gamma]}
+ \log (1+Y_{k-1}^{[-\gamma]} ) \star s_K
+ ({\rm source}),
\\[3mm]
\log \fba_k &= \log (1+\fba_k) \star K_f - \log (1+\fa_k) \star K_f^{[-2+2\gamma]}
+ \log (1+Y_{k-1}^{[+\gamma]} ) \star s_K
+ ({\rm source}),
\label{fa NLIE}
\end{align}
where the analyticity of $Y_{k-1} (v)$ for $-\gamma \le {\rm Im} \, v \le + \gamma$ is assumed, and the kernel $K_f$ is defined by 
\begin{equation}
K_f (v) = \frac{1}{2\pi i} \, \frac{\partial}{\partial v} \log S_f (v),\qquad
S_f (v) = \frac{\Gamma \( \frac{2-i v}{4} \) \Gamma \( \frac{i v}{4} \)}
{\Gamma \( -\frac{i v}{4} \) \Gamma \( \frac{2+i v}{4} \) } \,.
\end{equation}
In the limit $\gamma \to 0$ the kernels $\ts_K$ and $K_f^{[\pm 2]}$ have a pole at the origin, and we need the principal value prescription as in \eqref{principal value kernel}. Numerical computation is easier if we leave $\gamma > 0$ finite.
The source terms can be fixed by considering the asymptotic behavior ${\rm Re} \, v \to \pm \infty$ \cite{Su99}.

The two NLIE \eqref{fa NLIE} provides a closed set of equations for the minimal parameter list \eqref{parameters XXX1}. The product type relation \eqref{RecRec XXX} is a corollary of \eqref{fa NLIE},
\begin{equation}
\log \fa_k^{[-1+\gamma]} \, \fba_k^{[+1-\gamma]} = \log (1+Y_{k-1}).
\end{equation}
Cancellation of the whole source terms can be checked from the explicit results of \cite{Su99}.

\section{Hybrid NLIE from TQ-relations}\label{sec:exact 1s}

We will derive the hybrid NLIE starting from TQ-relations in the horizontal strips of the $\alg{su}(2|4|2)$-hook. The TQ-relations will play the same r\^ole as the covariant recursions for spinons in Section \ref{sec:NLIE XXX}.
Since we also use the TBA equations for the mirror \AdSxS\ in the $\sl(2)$ sector, we assume that all Y- and T-functions are invariant under the interchange $(a,s) \leftrightarrow (a,-s)$.\footnote{Recall that $T_{a,s}$ depend on gauge choice. It is helpful to introduce $\tau_a$ as in Appendix \ref{sec:Y and T} to discuss $T_{a,s>0}$ and $T_{a,s<0}$ in a symmetric way.}

\subsection{TQ-relation from Wronskian}\label{sec:TQ Wronskian}

First of all, we rederive the TQ-relations in the horizontal strip $s \ge 2$ of the $\alg{su}(2|4|2)$-hook, starting from the Wronskian formula of \cite{GKLT10}. The formula says that the T-functions $T_{a,s}$ for $a=0,1,2$ are given by
\begin{alignat}{7}
T_{0,s} &= {\sf Q}_{\overline{\emptyset}}^{[-s]} \,,& &
\notag \\[1mm]
T_{1,s} &= {\sf Q}_{1}^{[+s]} \, {\sf Q}_{\overline{1}}^{[-s]}
- {\sf Q}_{\overline{2}}^{[-s]} \, {\sf Q}_{2}^{[+s]} &\qquad &(s \ge 1),
\label{Wronskian T1s} \\[1mm]
T_{2,s} &= {\sf Q}_{12}^{[+s]} \, {\sf Q}_{\overline{12}}^{[-s]} &\qquad &(s \ge 2).
\notag
\end{alignat}
By solving the T-system
\begin{equation}
T_{a,s}^- \, T_{a,s}^+ = T_{a,s-1} \, T_{a,s+1} + T_{a-1,s} \, T_{a+1,s} \,,
\label{T system}
\end{equation}
at $(a=1,s \ge 2)$, we find
\begin{equation}
{\sf Q}_{12}^{[+s]} = {\rm det}
\begin{pmatrix}
{\sf Q}_{1}^{[s+1]} & {\sf Q}_{2}^{[s+1]} \\
{\sf Q}_{1}^{[s-1]} & {\sf Q}_{2}^{[s-1]} \\
\end{pmatrix},
\qquad
{\sf Q}_{\overline{\emptyset}}^{[-s]} \, {\sf Q}_{\overline{12}}^{[-s]} = {\rm det}
\begin{pmatrix}
{\sf Q}_{\overline{1}}^{[-s-1]} & {\sf Q}_{\overline{2}}^{[-s-1]} \\
{\sf Q}_{\overline{1}}^{[-s+1]} & {\sf Q}_{\overline{2}}^{[-s+1]}
\end{pmatrix}.
\label{Q12 det}
\end{equation}
Given \eqref{Wronskian T1s}, one can derive the relations
\begin{alignat}{7}
{\sf Q}_1^{[s-2]} \, T_{1,s} 
&- {\sf Q}_{1}^{[+s]} \, T_{1,s-1}^-
& &= {\sf Q}_{\overline{2}}^{[-s]} \, {\sf Q}_{12}^{[s-1]} \,,
\notag \\[1mm]
{\sf Q}_{\overline{1}}^{[-s+1]} \, T_{1,s}
&- {\sf Q}_{\overline{1}}^{[-s]} \, T_{1,s-1}^+
& &= {\sf Q}_{2}^{[+s]} \( {\sf Q}_{\overline{\emptyset}}^{[-s+1]} \, {\sf Q}_{\overline{12}}^{[-s+1]}\),
\notag \\[1mm]
{\sf Q}_{2}^{[s-2]} \, T_{1,s} 
&- {\sf Q}_{2}^{[+s]} \, T_{1,s-1}^-
& &= {\sf Q}_{\overline{1}}^{[-s]} \, {\sf Q}_{12}^{[s-1]} \,,
\notag \\[1mm]
{\sf Q}_{\overline{2}}^{[-s+2]} \, T_{1,s} 
&- {\sf Q}_{\overline{2}}^{[-s]} \, T_{1,s-1}^+
& &= {\sf Q}_{1}^{[+s]} \( {\sf Q}_{\overline{\emptyset}}^{[-s+1]} \, {\sf Q}_{\overline{12}}^{[-s+1]}\),
\label{Wronskian TQ rel}
\end{alignat}
for $s \ge 2$. They can be summarized as the TQ-relations of $A_1$ theory \cite{KLWZ96}:
\begin{alignat}{7}
Q_{1,s-1}^{\alpha\,-} \, T_{1,s} - Q_{1,s}^\alpha \, T_{1,s-1}^- &= \olQ_{1,s-1}^{\alpha\,-} \, L_{1,s} \,,
\notag \\[1mm]
\olQ_{1,s-1}^{\alpha\,+} \, T_{1,s} - \olQ_{1,s}^\alpha \, T_{1,s-1}^+ &= Q_{1,s-1}^{\alpha\,+} \, \olL_{1,s} \,,
\label{covariant TQ relation}
\end{alignat}
where
\begin{equation}
L_{1,s} = {\sf Q}_{12}^{[s-1]} \,, \qquad
\olL_{1,s} = {\sf Q}_{\overline{\emptyset}}^{[-s+1]} \, {\sf Q}_{\overline{12}}^{[-s+1]} \,,
\label{LL to sfQ}
\end{equation}
and $\alpha = {\rm I}, {\rm II}$ refers to
\begin{equation}
( Q_{1,s}^{\rm I} \,, \olQ_{1,s}^{\rm I} ) = ( {\sf Q}_{1}^{[+s]} \,, {\sf Q}_{\overline{2}}^{[-s]}),
\qquad
( Q_{1,s}^{\rm II} \,, \olQ_{1,s}^{\rm II} ) = ( {\sf Q}_{2}^{[+s]} \,, {\sf Q}_{\overline{1}}^{[-s]}).
\label{Q to sfQ}
\end{equation}
From \eqref{LL to sfQ} it follows that
\begin{equation}
T_{2,s} \, T_{0,s} = L_{1,s}^+ \, \olL_{1,s}^-
\qquad (s \ge 2).
\label{TT to LL}
\end{equation}

We wrote down the TQ-relations \eqref{covariant TQ relation} in a covariant way; the equations maintain the full gauge symmetry of the T-system, as long as $X_{1,s}$\,, {\it i.e.} any quantity with the lower index $(1,s)$, behaves in the same way as $T_{1,s}$\,. One can apply any gauge transformation $\Phi_{1,s}$ to them, and the equations remain invariant.
Since the Q-functions were originally translationally invariant $(X_{1,s} = X_{1,s-1}^+ \,,\ \olX_{1,s} = \olX_{1,s-1}^-)$\,, the new Q-functions are now translationally invariant modulo gauge transformation,
\begin{equation}
\frac{L_{1,s} }{L_{1,s-1}^+ } = \frac{Q_{1,s}^\alpha }{Q_{1,s-1}^{\alpha \,+} } = \frac{\Phi_{1,s} }{\Phi_{1,s-1}^+ } \,,
\qquad
\frac{\olL_{1,s} }{\olL_{1,s-1}^- } = \frac{\olQ_{1,s}^\alpha }{\olQ_{1,s-1}^{\alpha \,-} } = \frac{\Phi_{1,s} }{\Phi_{1,s-1}^- } \,,
\qquad
\frac{\Phi_{1,s}^- \, \Phi_{1,s}^+ }{\Phi_{1,s-1} \,, \Phi_{1,s+1} } = 1.
\label{translation QL}
\end{equation}
These equations are valid for $s \ge 3$, because $(L_{1,s} \,, \olL_{1,s})$ are not defined at $s=1$.
The last equation of \eqref{translation QL} means that $\Phi$ is a gauge degree of freedom.

\bigskip
Now consider the inverse problem, that is to obtain Q-functions when $T_{1,s}$ are given for a certain range of $s$.
If $(L_{1,s} \,, \olL_{1,s})$ are also given, it reduces to the problem to solve some second-order difference equations \cite{KLWZ96}. To see this, we rearrange the first line (resp. the second line) of \eqref{covariant TQ relation} into a difference equation for $Q$'s (resp. $\olQ$'s) as
\begin{align}
\frac{L_{1,s}^{[-2]} \, Q_{1,s} }{\Phi_{1,s}^{[-2]} }
+ \frac{L_{1,s} \, Q_{1,s}^{[-4]} }{\Phi_{1,s}^{[-4]} }
- \frac{R_s \, Q_{1,s}^{[-2]} }{\Phi_{1,s}^{[-2]} } = 0,
\qquad
\frac{\olL_{1,s}^{[+2]} \, \olQ_{1,s} }{\Phi_{1,s}^{[+2]} }
+ \frac{\olL_{1,s} \, \olQ_{1,s}^{[+4]} }{\Phi_{1,s}^{[+4]} }
- \frac{\olR_s \, \olQ_{1,s}^{[+2]} }{\Phi_{1,s}^{[+2]} } = 0,
\label{covariant Q difference}
\end{align}
where
\begin{align}
R_s = \frac{\Phi_{1,s-1}^-}{T_{1,s-1}^- } \(
\frac{L_{1,s} \, T_{1,s-2}^{[-2]} }{\Phi_{1,s-2}^{[-2]} }
+ \frac{L_{1,s}^{[-2]} \, T_{1,s} }{\Phi_{1,s}^{[-2]} } \),
\qquad
\olR_s = \frac{\Phi_{s-1}^+}{T_{1,s-1}^+ } \(
\frac{\olL_{1,s} \, T_{1,s-2}^{[+2]} }{\Phi_{s-2}^{[+2]} }
+ \frac{\olL_{1,s}^{[+2]} \, T_{1,s} }{\Phi_{1,s}^{[+2]} } \).
\end{align}
In general, the second order difference equations have two linearly independent solutions, in agreement with \eqref{Q to sfQ}.

\bigskip
In the spectral problem of \AdSxS, we do not try to solve the difference equations \eqref{covariant Q difference}, because we do not know the exact form of $(L_{1,s} \,, \olL_{1,s})$. Instead, we will construct a set of hybrid NLIE from the covariant TQ-relations \eqref{covariant TQ relation}. To formulate hybrid NLIE, we use the analyticity data of the asymptotic solutions, which will be discussed in Appendix \ref{app:asymp Wronskian}.

\subsection{Decomposing TBA from TQ-relations}

In this subsection, we decompose the TBA equations (or the Y-system) of $A_1$ theory using TQ-relations. Just like \eqref{relation AQL XXX}, we introduce the `spinon' variables by
\begin{equation}
A_{1,s}^\alpha = \frac{\olQ_{1,s-1}^{\alpha \,-} }{Q_{1,s-1}^{\alpha \,-} } \, L_{1,s} \,,
\qquad
\olA_{1,s}^\alpha = \frac{Q_{1,s-1}^{\alpha \,+} }{\olQ_{1,s-1}^{\alpha \,+} } \, \olL_{1,s} \,,
\qquad (s \ge 2, \ \alpha = {\rm I}, {\rm II}),
\label{relation AQL}
\end{equation}
and simplify the covariant $A_1$ TQ-relations \eqref{covariant TQ relation} as
\begin{equation}
T_{1,s}^+ - A_{1,s}^{\alpha \,+} = \frac{\olA_{1,s}^{\alpha \,+} }{\olA_{1,s-1}^\alpha } \, T_{1,s-1} \,,
\qquad
T_{1,s}^- - \olA_{1,s}^{\alpha \,-} = \frac{A_{1,s}^{\alpha \,-} }{A_{1,s-1}^\alpha } \, T_{1,s-1} \,,
\qquad (s \ge 3).
\label{TA rec}
\end{equation}
The lower bound of $s$ has increased by one, because $(L_{1,s} \,, \olL_{1,s})$ are defined only for $s \ge 2$.

Following Section \ref{sec:decomp XXX}, let us define new gauge-invariant variables by
\begin{equation}
1 + \fb_s^\alpha = \frac{T_{1,s}^+ }{A_{1,s}^{\alpha \,+} } 
= \frac{Q_{1,s-1}^\alpha }{\olQ_{1,s-1}^\alpha } \, \frac{T_{1,s}^+ }{L_{1,s}^+ } \,,
\qquad
1 + \fbb_s^\alpha = \frac{T_{1,s}^- }{\olA_{1,s}^{\alpha \,-} }
= \frac{\olQ_{1,s-1}^\alpha }{Q_{1,s-1}^\alpha } \, \frac{T_{1,s}^- }{\olL_{1,s}^- } \,,
\qquad (s \ge 2),
\label{def:fbb 1s}
\end{equation}
and count the number of variables and the number of equations. The Y-functions $Y_{1,M \ge s}$ are now replaced by
\begin{equation}
(1 + \fb_M^\alpha) \, (1 + \fbb_M^\alpha) = \frac{T_{1,M}^- \, T_{1,M}^+}{T_{0,M} \, T_{2,M} } = 1+Y_{1,M} \,, \qquad (\alpha = {\rm I}, {\rm II}).
\label{1+b to Y1s}
\end{equation}
This procedure corresponds to the middle of Figure \ref{fig:Truncate}, where the TBA equations are decomposed into new degrees of freedom.

Unfortunately, as discussed in Section \ref{sec:decomp XXX}, there is always one more unknown variables than the number of the equations that can be derived algebraically. For example, one can consider the general parameter list
\begin{equation}
\pare{Y_{1,2} \,, \dots Y_{1,s-1} } \ \bigcup \ 
\pare{ \begin{array}{ccccc}
\fb_s^\alpha & \fb_{s+1}^\alpha & \dots & \fb_{s+\ell}^\alpha \\[1mm]
\fbb_s^\alpha & \fbb_{s+1}^\alpha & \dots & \fbb_{s+\ell}^\alpha
\end{array} },
\qquad (s \ge 3),
\label{parameter list ell}
\end{equation}
where $\alpha$ is either ${\rm I}$ or ${\rm II}$. Just like \eqref{parameters XXX3}, the equations cannot be closed as long as $\ell$ is finite. In other words, the equations can be closed if in the limit $\ell \to \infty$ we require that $\lim_{s \to \infty} \( \fb_s \,, \fbb_s \)$ approach the asymptotic functions.\footnote{See \cite{Gromov09,BH10a} for the discussion on boundary conditions of $Y_{a,s}$ as $a \to \infty$ or $s \to \infty$.}
This conclusion is unchanged even if one includes both $\alpha={\rm I}, {\rm II}$ in the above parameter list.

The shortage of one equation can be understood as follows. As we saw in Section \ref{sec:TQ Wronskian}, the most general solution of T-system or TQ-relations is given by fundamental Q-functions. However, it also implies that there is no equation which determines the fundamental Q-functions in an algebraic manner.
To supply more constraints, we need to study analyticity conditions.

\subsection{Analyticity conditions in the horizontal strip}\label{sec:ANZC 1s}

We repeat the discussion in Section \ref{sec:ANZC XXX} in the case of the horizontal part of the $\alg{su}(2|4|2)$-hook. Although the asymptotic Q-functions on the $\alg{su}(2|4|2)$-hook are different from those of the XXX model, both of them have good analytic properties on the upper or lower half of the complex rapidity plane.

From \eqref{TA rec}, we can derive the relation
\begin{align}
\fb_s^{\alpha} &= \frac{\olA_{1,s}^{\alpha \,+} }{A_{1,s}^{\alpha \,+} } \, \frac{T_{1,s-1} }{\olA_{1,s-1}^{\alpha} }
= \frac{Q_{1,s-1}^{\alpha \,[+2]} }{\olQ_{1,s-1}^{\alpha} } \, \frac{T_{1,s-1} }{L_{1,s-1}^{[+2]} } \,,
\label{B to AA 1s-1} \\[1mm]
\fbb_s^{\alpha} &= \frac{A_{1,s}^{\alpha \,-} }{\olA_{1,s}^{\alpha \,-} } \, \frac{T_{1,s-1} }{A_{1,s-1}^{\alpha} }
= \frac{\olQ_{1,s-1}^{\alpha \,[-2]} }{Q_{1,s-1}^{\alpha} } \, \frac{T_{1,s-1} }{\olL_{1,s-1}^{[-2]} } \,,
\label{B to AA 1s-2}
\end{align}
for $s \ge 3$. We take the Fourier transform of the logarithmic derivative. As for the left hand side, we introduce the regularization
\begin{equation}
\fa_s^\alpha (v) = \fb_s^\alpha \(v-\frac{i \gamma}{g}\), \qquad
\fba_s^\alpha (v) = \fbb_s^\alpha \(v+\frac{i \gamma}{g}\), \qquad
0 < \gamma < 1,
\label{def:fa s}
\end{equation}
and take the Fourier transform on the real axis of $(\fa_s(v), \fba_s(v))$. The upper bound of $\gamma$ comes from the definition of $\fb$'s in \eqref{def:fbb 1s}; for instance, the variable $L_{1,s}^{[+1-\gamma]}$ should stay on the upper half plane. As for the right hand side, we assume that each factor in \eqref{B to AA 1s-1} is analytic inside the strip $-\frac{\gamma}{g} \le {\rm Im} \, v \le 0$, and each factor in \eqref{B to AA 1s-2} is analytic inside the strip $0 \le {\rm Im} \, v \le + \frac{\gamma}{g}$\,.\footnote{Again, these assumptions are part of the analyticity conditions we will use below.}
Then we obtain, for $s \ge 3$,
\begin{align}
\hdl \fa_s^{\alpha} &= e^{- \gamma \frac{q}{g}} \pare{ \hdl Q_{1,s-1}^{\alpha \,[+2]} - \hdl \olQ_{1,s-1}^{\alpha} - \hdl L_{1,s-1}^{[+2]} + \hdl T_{1,s-1} },
\notag \\[1mm]
\hdl \fba_s^{\alpha} &= e^{+ \gamma \frac{q}{g}} \pare{ \hdl \olQ_{1,s-1}^{\alpha \,[-2]} - \hdl Q_{1,s-1}^{\alpha} - \hdl \olL_{1,s-1}^{[-2]} + \hdl T_{1,s-1} }.
\label{bs Fourier}
\end{align}
Note that the terms $\hdl Q_{1,s-1}^{\alpha \,[+2]},\ \hdl \olQ_{1,s-1}^{\alpha \,[-2]}$ are again as dangerous as in \eqref{danger Qj}.

\subsubsection*{ANZC conditions}

We assume that the analyticity data of the relevant functions are same as in the asymptotic case. The asymptotic expressions for $T_{1,s} \,, L_{1,s} \,, \olL_{1,s}$ are discussed in Appendix \ref{app:asymp sol}.
We use the gauge $T_{1,s} = \TH_{1,s}$ and set $\Phi_{1,s}=1$.\footnote{This is equivalent to $\Theta_{1,s}=1$ in the notation of Appendix \ref{app:asymp sol}.}

Suppose that for $s \ge 2$,
\begin{equation}
T_{1,s} (v) \quad \text{is  ANZC  for} \quad -\frac{1}{g} \le \,{\rm Im} \, v \le \frac{1}{g} \,.
\label{analytic T1s-1}
\end{equation}
Following the same argument as before, and keeping in mind the definition of $\fb$'s in \eqref{def:fbb 1s} and $f^\pm$ in \eqref{def:fmpm}, we obtain
\begin{align}
0 &= e^{+\frac{q}{g} } \hdl T_{1,s}^- - e^{-\frac{q}{g} } \hdl T_{1,s}^+ ,
\notag \\[1mm]
&= e^{+\frac{q}{g} } \hdl \pare{ (1+\fbb_s^\alpha) \, \olL_{1,s}^- \, \frac{Q_{1,s-1}^\alpha }{\olQ_{1,s-1}^\alpha } }
- e^{-\frac{q}{g} } \hdl \pare{ (1+\fb_s^\alpha) \, L_{1,s}^+ \, \frac{\olQ_{1,s-1}^\alpha }{Q_{1,s-1}^\alpha } } \,.
\end{align}
If the Fourier transform of the logarithmic derivative of each factor is well-defined, we get
\begin{align}
0 &= e^{+\frac{q}{g} } \hdl (1+\fbb_s^\alpha) - e^{-\frac{q}{g} } (1+\fb_s^\alpha)
\notag \\[1mm]
&\qquad + 2 \ssp \cosh \( \frac{q}{g} \) \( \hdl Q_{1,s-1}^\alpha - \hdl \olQ_{1,s-1}^\alpha \)
+ e^{+\frac{q}{g} } \hdl \olL_{1,s}^- - e^{-\frac{q}{g} } \hdl L_{1,s}^+ \,.
\label{TmTp analytic}
\end{align}
We also assume that for $s \ge 3$,
\begin{alignat}{9}
Q_{1,s-1}^\alpha (v) \ \ &\text{is ANZC for} \ \ {\rm Im} \, v \ge 0,
&\qquad
\olQ_{1,s-1}^\alpha (v) \ \ &\text{is ANZC for} \ \ {\rm Im} \, v \le 0,
\notag \\[2mm]
L_{1,s-1} (v) \ \ &\text{is ANZC for} \ \ {\rm Im} \, v \ge +\frac{1}{g} \,,
&\qquad
\olL_{1,s-1} (v) \ \ &\text{is ANZC for} \ \ {\rm Im} \, v \le -\frac{1}{g} \,.
\label{analytic cQL}
\end{alignat}
As corollaries, it follows that
\begin{alignat}{9}
\hdl Q_{1,s-1}^\alpha (q) &= 0 \quad {\rm for} \ \ {\rm Re} \; q > 0,
&\qquad
\hdl \olQ_{1,s-1}^\alpha (q) &= 0 \quad {\rm for} \ \ {\rm Re} \; q < 0,
\notag \\[2mm]
\hdl L_{1,s-1}^+ (q) &= 0 \quad {\rm for} \ \ {\rm Re} \; q > 0,
&\qquad
\hdl \olL_{1,s-1}^- (q) &= 0 \quad {\rm for} \ \ {\rm Re} \; q < 0.
\end{alignat}
By using the analyticity and the translational properties \eqref{translation QL} for the gauge $\Phi_{1,s}=1$, we find
\begin{equation}
e^{-\frac{q}{g} } \hdl L_{1,s}^+ = e^{-\frac{q}{g} } \hdl L_{1,s-1}^{[+2]} = \hdl L_{1,s-1}^+ \,,
\quad
e^{+\frac{q}{g} } \hdl \olL_{1,s}^- = e^{+\frac{q}{g} } \hdl \olL_{1,s-1}^{[-2]} = \hdl \olL_{1,s-1}^- \,.
\end{equation}
The equation \eqref{TmTp analytic} leads to
\begin{alignat}{5}
\hdl Q_{1,s-1}^\alpha - \hdl L_{1,s-1}^+ \, \hat s_K (q) &= 
\begin{cases}
0 &\ \ {\rm Re} \; q > 0, \\
\Big( - e^{+\frac{q}{g} } \hdl (1+\fbb_s^\alpha) + e^{-\frac{q}{g} } \hdl (1+\fb_s^\alpha) \Big) \, \hat s_K (q) &\ \ {\rm Re} \; q < 0,
\end{cases}
\notag \\[2mm]
\hdl \olQ_{1,s-1}^\alpha - \hdl \olL_{1,s-1}^- \, \hat s_K (q) &=
\begin{cases}
\Big( - e^{-\frac{q}{g} } \hdl (1+\fb_s^\alpha) + e^{+\frac{q}{g} } \hdl (1+\fbb_s^\alpha) \Big) \, \hat s_K (q) &\ \ {\rm Re} \; q > 0, \\
0 &\ \ {\rm Re} \; q < 0.
\end{cases}
\label{cQQ to Lambda}
\end{alignat}

We employ the regularization \eqref{def:fa s} also for $(1+\fb_s) \,, (1+\fbb_s)$, assuming that
\begin{alignat}{7}
1+\fb_s (v) \ \ &\text{is ANZC for} &\ \ -\frac{\gamma}{g} \le \, &{\rm Im} \, v \le 0,
\notag \\[1mm]
1+\fbb_s (v) \ \ &\text{is ANZC for} &\ \ 0 \le \, &{\rm Im} \, v \le +\frac{\gamma}{g}\,.
\label{analytic 1+bs}
\end{alignat}
It follows that
\begin{equation}
\hdl (1+\fb_s) = e^{+ \gamma \frac{q}{g}} \hdl (1+\fa_s), \qquad
\hdl (1+\fbb_s) = e^{- \gamma \frac{q}{g}} \hdl (1+\fba_s).
\label{def:fas Fourier}
\end{equation}

\subsubsection*{Derivation of NLIE}

Let us rewrite T-function into Y-function \eqref{1+b to Y1s} using analyticity \eqref{analytic T1s-1},
\begin{equation}
\hdl T_{1,s-1} = \( \hdl L_{1,s-1}^+ + \hdl \olL_{1,s-1}^- + \hdl (1+Y_{1,s-1}) \) \hat s_K (q),
\qquad s \ge 3.
\end{equation}
The equation \eqref{bs Fourier} then simplifies with the help of \eqref{analytic cQL}, and we obtain
\begin{align}
e^{+\gamma \frac{q}{g}} \hdl \fa_s^{\alpha} &= e^{+\frac{2q}{g}} \[ \hdl Q_{1,s-1}^{\alpha} - \hdl L_{1,s-1}^+ \, \hat s_K (q) \]
\notag \\[1mm]
&\hspace{20mm} - \[ \hdl \olQ_{1,s-1}^{\alpha} - \hdl \olL_{1,s-1}^- \, \hat s_K (q) \]
+ \hdl (1+Y_{1,s-1}) \, \hat s_K (q),
\notag \\[1mm]
e^{-\gamma \frac{q}{g}} \hdl \fba_s^{\alpha} &= e^{-\frac{2q}{g}} \[ \hdl \olQ_{1,s-1}^{\alpha} - \hdl \olL_{1,s-1}^- \, \hat s_K (q) \]
\notag \\[1mm]
&\hspace{20mm} - \[ \hdl Q_{1,s-1}^{\alpha} - \hdl L_{1,s-1}^+ \, \hat s_K (q) \]
+ \hdl (1+Y_{1,s-1}) \, \hat s_K (q).
\label{bs Fourier2}
\end{align}
We substitute \eqref{cQQ to Lambda} to these equations, keeping in mind that the terms in the square brackets are identical to the combination we found in \eqref{cQQ to Lambda}. Recalling also \eqref{def:fa Fourier}, we find
\begin{align}
\hdl \fa_s^\alpha &= \frac{e^{-\abs{\frac{q}{g} }} }{2 \ssp \cosh \frac{q}{g} } \, \hdl (1+\fa_s^\alpha)
- \frac{e^{+2\frac{q}{g} (1-\gamma) - \abs{\frac{q}{g} }} }{2 \ssp \cosh \frac{q}{g} } \, \hdl (1+\fba_s^\alpha)
+ \frac{e^{-\gamma \frac{q}{g} } }{2 \ssp \cosh \frac{q}{g} } \, \hdl (1+Y_{1,s-1}) ,
\notag \\[2mm]
\hdl \fba_s^\alpha &= \frac{e^{-\abs{\frac{q}{g} }} }{2 \ssp \cosh \frac{q}{g} } \, \hdl (1+\fba_s^\alpha)
- \frac{e^{-2\frac{q}{g} (1-\gamma) - \abs{\frac{q}{g} }} }{2 \ssp \cosh \frac{q}{g} } \, \hdl (1+\fa_s^\alpha)
+ \frac{e^{-\gamma \frac{q}{g} } }{2 \ssp \cosh \frac{q}{g} } \, \hdl (1+Y_{1,s-1}) .
\label{mm Fourier3}
\end{align}
The inverse Fourier transform gives
\begin{align}
\log \fa_s^\alpha &= \log (1+\fa_s^\alpha) \star K_f
- \log (1+\fba_s^\alpha) \star K_f^{[+2-2\gamma]}
+ \log (1+Y_{1,s-1}^{[-\gamma]} ) \star s_K + ({\rm source}),
\notag \\[2mm]
\log \fba_s^\alpha &= \log (1+\fba_s^\alpha) \star K_f 
- \log (1+\fa_s^\alpha) \star K_f^{[-2+2\gamma]}
+ \log (1+Y_{1,s-1}^{[+\gamma]} ) \star s_K + ({\rm source}),
\label{a NLIE}
\end{align}
where the kernel $K_f$ is now defined by \eqref{def:Kf}, and the source terms can be fixed by analyticity data, as done in Introduction.

\bigskip
In summary, the minimal parameter list for the horizontal strips of the $\alg{su}(2|4|2)$-hook is
\begin{equation}
\pare{Y_{1|w} \,, \dots \,, Y_{s-2|w} } \ \bigcup \ 
\pare{ \begin{array}{cccc}
\fa_s \\[1mm]
\fba_s
\end{array} }, \qquad (s \ge 3).
\label{parameters min s}
\end{equation}
We cannot remove $Y_{1|w}$\,, because the pair $(\fa_2 \,, \fba_2)$ are not related to $Y_{1,1}=-1/Y_-$ in \eqref{bs Fourier}.\footnote{In fact, we can find $\frac{T_{1,1}^+ \, T_{1,1}^-}{L_{1,1}^+ \, \olL_{1,1}^-} = \frac{1-1/Y_-}{1-1/Y_+}$\,, if we use the asymptotic formulae of $L_{1,s} \,, \olL_{1,s}$ at $s=1$.}
To determine $\pare{Y_{1|w} \,, \dots \,, Y_{s-2|w} }$, we use the simplified TBA for $Y_{M|w} \ (M=1, \dots, s-3)$,
\begin{equation}
\log Y_{1,M+1} = \log(1+Y_{1,M})(1+Y_{1,M+2})\star s_K
+ \delta_{M,1}\, \log \frac{1-\frac{1}{Y_-} }{1-\frac{1}{Y_+} } \hstar s_K \,,
\label{simp TBA YMw}
\end{equation}
with $Y_{0|w}=0$, and as for $Y_{s-2|w}$
\begin{equation}
\log Y_{s-2|w} = \log(1+Y_{s-3|w}) \, (1+\fa_s^{\alpha \, [+\gamma]} ) \, (1+\fba_s^{\alpha \, [-\gamma]}) \star s_K
+ \delta_{s-2,1}\, \log \frac{1-\frac{1}{Y_-} }{1-\frac{1}{Y_+} } \hstar s_K \,.
\label{simp TBA Ys-1w}
\end{equation}
The NLIE \eqref{a NLIE} are used to determine $(\fa_s \,, \fba_s)$.

\section{Discussion}\label{sec:discussion}

In this paper, we derived hybrid NLIE from two setups. The first setup was a pair of covariant recursions for spinons, and the second setup was $A_1$ TQ-relations. The $A_1$ TQ-relations appeared in the horizontal strips of the $\alg{su}(2|4|2)$-hook.
By combining TBA equations and the equations for auxiliary variables, we replaced the TBA equations for $Y_{M|w}$ for $M \ge 2$ by a couple of auxiliary variables. We used the assumptions \eqref{analytic T1s-1}, \eqref{analytic cQL} and \eqref{analytic 1+bs}.

The hybrid NLIE provides us with an efficient algorithm to compute the exact spectrum of \AdSxS\ string theory, both numerically and analytically. It is desirable to develop similar techniques to truncate the vertical strip $a \ge 2$ of the $\alg{su}(2|4|2)$-hook, by generalizing TQ-relations and hybrid NLIE for a higher-rank system \cite{KLWZ96,DK06,Damerau10}.
In the horizontal strip, the analyticity data were relatively simple, and the source terms in NLIE have simple structure. In the vertical strip, however, the analytic structure would be much complicated.

It is interesting to clarify the physical and mathematical interpretation of new auxiliary degrees of freedom.
The relation \eqref{Y to pp Kon} suggests that the $M|w$-strings appearing in the string hypothesis of the mirror \AdSxS\ are boundstates of two fundamental excitations; like mesons and quarks.
Moreover, the $\alg{su}(2|4|2)$-hook is replaced by another hook as in Figure \ref{fig:Truncate}, which would have deeper relation with representation theory of $\alg{su}(2|4|2)$, like \cite{Su98}.
Such observation might give us a hint in searching for hidden structure of the mirror \AdSxS\ theory, along the line of \cite{FR86,DdV87,JMS06,JMS08a,JMS08b,JMS09,JMS10}.

\subsubsection*{Note added in v4}

After the submission to arXiv, we are reminded of the talk \cite{Gromov10conf} in Stockholm,
where another way to truncate the horizontal wings was announced.
The claim is as follows. By choosing a suitable gauge, one can assume that the exact T-function $T_{1,s}$ is given by the ansatz\footnote{Reasoning for this ansatz has recently been explained in \cite{GKLV11}. The author thanks N. Gromov, S. Leurent and D. Volin for discussions about this method.}
\begin{equation}
T_{1,s} (v) = s + \int_{-\infty}^{+\infty} dt \; K_s (v-t) f (t), \qquad (s \ge 1).
\label{Gromov ansatz}
\end{equation}
The variable $f$ is the density responsible for the discontinuities of $T_{1,s}$\,, and this unknown variable is determined by solving\footnote{Here the definition of $f^\pm$ can be different from \eqref{def:fmpm}.}
\begin{equation}
\frac{1+Y_{1,1}}{1+1/Y_{2,2}} = \frac{T_{1,1}^- \, T_{1,1}^+ \, T_{2,3}}{T_{2,2}^- \, T_{2,2}^+ \, T_{0,1} }
= \frac{(1+K_1^+ \star_{p.v.} f + f/2) \, (1+K_1^- \star_{p.v.} f + f/2)}{(1+K_1^+ \star_{p.v.} f - f/2) \, (1+K_1^- \star_{p.v.} f - f/2)} \,.
\label{Gromov NLIE}
\end{equation}
%The derivation of this equation has not been published anywhere. 
The equation \eqref{Gromov NLIE}, allows us to truncate the horizontal strips of the $\alg{su}(2|4|2)$-hook, and this result is claimed to be consistent with the numerical data for Konishi state for $0 \lesssim \lambda \lesssim 1000$.

\subsubsection*{Acknowledgements}

The author acknowledges Sergey Frolov, Gleb Arutyunov and Zoltan Bajnok for helpful discussions throughout this project. The author thanks Junji Suzuki and Andreas Kl\"umper for comments on the draft. He also thanks to the organizers of {\it Workshop on Integrable Quantum Systems} for an inspiring talk of \cite{Klumper talk}.
This work is support by the Netherlands Organization for Scientific Research (NWO) under the VICI grant 680-47-602.

\appendix

\section{Notations}\label{app:notations}

We use the notation
\begin{gather}
g = \frac{\sqrt \lambda}{2 \pi} \,, \qquad
f (v)^{[\pm m]} \equiv f \Big( v \pm \frac{im}{g} \mp i0 \Big), \qquad
f(v)^\pm = f (v)^{[\pm 1]},
\label{def:fmpm} \\[2mm]
x_s (v) = \frac{v}{2} \( 1 + \sqrt{ 1 - \frac{4}{v^2} } \, \),\qquad
x (v) = \frac12 \( v - i \sqrt{4 - v^2} \, \).
\end{gather}
Let us define
\begin{equation}
\cR_{(\pm)} (v) = \prod_{j=1}^K \frac{x(v) - x^\pm_{s,j}}{\sqrt{x^\pm_{s,j}}} \,,\quad
\cB_{(\pm)} (v) = \prod_{j=1}^K \frac{\frac{1}{x(v)} - x^\pm_{s,j}}{\sqrt{x^\pm_{s,j}}} \,,\quad
\cQ (v) = \prod_{j=1}^K (v-u_j) \,,
\label{def:RBQ}
\end{equation}
where $K$ is the number of physical excitations and $x^\pm_{s,j} = x_s (u_j \pm \frac{i}{g})$. The rapidity $u_j$ sits in the physical region of string theory.\footnote{Our definition of $(\pm)$ is different from \cite{GKV09a,GKKV09,GKV09b}.}
There are identities
\begin{equation}
\cR^{[m]}_{(\pm)} (v) \, \cB^{[m]}_{(\pm)} (v) = (-1)^K \, \cQ \Big( v + \frac{i}{g} (m \mp Q_j) \Big), \qquad
\cR^+_{(+)} \, \cB^+_{(+)} = \cR^-_{(-)} \, \cB^-_{(-)} = (-1)^K \, \cQ.
\label{RB into Q}
\end{equation}
Since $x(v) = x_s (v)$ for ${\rm Im} \, v < 0$ and $1/x (v) = x_s (v)$ for ${\rm Im} \, v > 0$, we observe that
\begin{equation}
\cR_{(\pm)} (v) \ \text{have no zeroes for} \ {\rm Im} \, v  > 0, \qquad
\cB_{(\pm)} (v) \ \text{have no zeroes for} \ {\rm Im} \, v  < 0.
\label{analyticity RB}
\end{equation}
The zeroes of $\cQ (v)$ lie on the real axis of $v$ if there are no boundstates.
The branch cuts of $\cR, \cB$ lie along the real axis of $v$.

We use the kernel
\begin{gather}
s_K (v) = \frac{g}{4 \cosh {\pi g v \ov 2}} \,,\qquad
\tilde s_K = s_K^- = - s_K^+ \,,
\label{def:sK} \\[1mm]
K_f (v) = \frac{1}{2\pi i} \, \frac{\partial}{\partial v} \log S_f (v),\qquad
S_f (v) = \frac{\raisebox{3.5mm}{$\ds \Gamma \( \frac{g}{4 \ssp i} \, \Big( v+\frac{2i}{g} \Big) \) \, \Gamma \( - \frac{g \ssp v}{4 \ssp i} \)$} }
{\raisebox{-3.5mm}{$\ds \Gamma \( \ssp \frac{g \ssp v}{4 \ssp i} \) \, \Gamma \( - \frac{g}{4 \ssp i} \, \Big( v-\frac{2i}{g} \Big) \)$} } \,,
\label{def:Kf}
\end{gather}
and the notation
\begin{equation}
F \star K (v) = \int_{-\infty}^\infty dt \, F(t) \, K (v-t),
\qquad
F \hstar K (v) = \int_{-2}^2 dt \, F(t) \, K (v-t).
\label{def:star}
\end{equation}
Note that all kernels in \eqref{def:Kf} are symmetric, $K(v)=K(-v)$. The asymptotic behavior of $S_f (v)$ is $S_f (v) \to \mp i$ as $v \to \pm \infty$.

As for the convolution with $\ts_K$ and $K_f^{[\pm 2]}$\,, one may use the principal-value prescription:
\begin{align}
\log f^\mp \star s_K &= \frac12 \, \log f \mp \log f \star_{p.v.} \ts_K \,,
\notag \\[2mm]
\log f \star K_f^{[+2 \pm 0]} &= \mp \frac12 \, \log f + \log f \star_{p.v.} K_f^{[+2]} \,,
\notag \\[2mm]
\log f \star K_f^{[-2 \pm 0]} &= \pm \frac12 \, \log f + \log f \star_{p.v.} K_f^{[-2]} \,.
\label{principal value kernel}
\end{align}

\section{Review of Y-system and T-system}\label{sec:Y and T}

\begin{figure}[htbp]
\begin{center}
\includegraphics[scale=0.7]{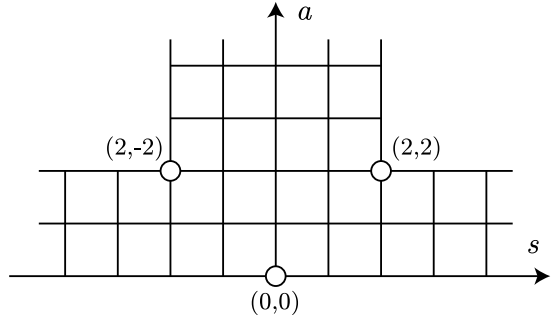}
\caption{The $\alg{su}(2|4|2)$-hook, whose boundaries lie along $(0,s), \, (2, s),\, (a, \pm 2)$.}
\label{fig:Thook}
\end{center}
\end{figure}

Most of the mirror TBA equations on \AdSxS, except for the exact energy and the exact Bethe roots, can be regarded as the Y-system on $\alg{su}(2|4|2)$ supplemented by certain analyticity conditions. The canonical definition of the $\alg{su}(2|4|2)$ Y-system is
\begin{equation}
Y_{a,s}^+ Y_{a,s}^- = \frac{\( 1+Y_{a,s+1} \) \( 1+Y_{a,s-1} \)}{\( 1 + \frac{1}{Y_{a+1,s}} \) \( 1 + \frac{1}{Y_{a-1,s}} \)} \,, \qquad
v \in (-2,2),
\label{canonical Y}
\end{equation}
where $(a,s)$ runs through the lattice points shown in Figure \ref{fig:Thook}.

\medskip

We assume the interchange symmetry $(a,s) \leftrightarrow (a,-s)$ in what follows. The Y-functions $Y_{a,s}$ are defined by
\begin{alignat}{9}
Y_{1,1} &\leftrightarrow - \frac{1}{Y_-} \,, &\quad
Y_{2,2} &\leftrightarrow - Y_+ \,, &\quad
Y_{M+1,1} &\leftrightarrow \frac{1}{Y_{M|vw}} \,,&\quad
Y_{1,M+1} &\leftrightarrow Y_{M|w} \,,
\notag \\
Y_{Q,0} &\leftrightarrow Y_Q \,, & & & & & &
\label{relate Y with Yas}
\end{alignat}
where we set chemical potentials to zero. We introduce the T-functions by
\begin{gather}
Y_{a,s} \equiv
\begin{cases}
\ds \tau_a \, \frac{T_{a,1} \, T_{a,-1}}{T_{a+1,0} \, T_{a-1,0}} &\ds \quad (s=0) \qquad {\rm where} \ \ \frac{\tau_a^+ \, \tau_a^-}{\tau_{a+1} \, \tau_{a-1}} = 1,
\\[5mm]
\ds \ \frac{T_{a,s+1} \, T_{a,s-1}}{T_{a+1,s} \, T_{a-1,s}} &\quad (s \neq 0).
\end{cases}
\label{def:Y to T}
\end{gather}
The canonical Y-system \eqref{canonical Y} are then solved by the T-system which live on the $\alg{su}(2|4|2)$-hook,
\begin{equation}
T_{a,s}^+ \, T_{a,s}^- = T_{a+1,s} \, T_{a-1,s} + T_{a,s+1} \, T_{a,s-1} \,,
\qquad (a \ge 1,\ s \neq 0),
\label{T system 1}
\end{equation}
and the Y-system at $\abs{s}=0,1$ are solved by
\begin{equation}
T_{a,0}^+ \, T_{a,0}^- = T_{a+1,0} \, T_{a-1,0} + \tau_a \, T_{a,1} \, T_{a,-1} \,,
\qquad (a \ge 1).
\label{T system 0}
\end{equation}
we introduced an extra `gauge' factor $\tau_a$ in order to take the asymptotic limit $\tau_a \to 0$ easily. By redefinition of $T_{a,s} \ (s \neq 0)$ one can recover the standard definition of T-system, as will be discussed later.

T-functions vanish outside the $\alg{su}(2|4|2)$-hook, $T_{-1,s} = T_{3, \pm Q} = T_{Q, \pm 3} = 0$ for $s \in \bb{Z},\ Q \ge 3$. Along the boundary of the hook, the T-system reduces to the discrete Laplace equation without source term.

The T-system equations \eqref{T system 1}, \eqref{T system 0} are invariant under the gauge transformation
\begin{equation}
T_{a,s} \ \to \ g_1^{[a+s]} \, g_2^{[a-s]} \, g_3^{[-a+s]} \, g_4^{[-a-s]} \, T_{a,s} \,,
\label{Tas gauge trs}
\end{equation}
with $\tau_a$ left intact. We need two gauge degrees of freedom to impose the boundary conditions $T_{0,s}=1$ for all $s$.

There are subtle points in the study of the exact spectrum from the Y- and T-system on the $\alg{su}(2|4|2)$-hook, compared to the study from the mirror TBA equations on \AdSxS.
Firstly, The Y-system at the corner $(a,s)=(2,\pm 2)$ does not follow directly from the mirror TBA. 
To derive them, we have to use the parametrization \eqref{def:Y to T} and the T-system \eqref{T system 1} except at the corner. Then, the T- and Y-system at the corner can be derived from the equations in the neighborhood.
Secondly, if the mirror rapidity $v$ lies $(-\infty,-2) \cup (2, \infty)$, the Y-system is no longer canonical, and we need to know the gap on branch cuts.\footnote{The Y-system remains canonical if we use $f^\pm = f (v \pm \frac{i}{g} + i0)$ or $f^\pm = f (v \pm \frac{i}{g} - i0)$ instead of \eqref{def:fmpm} \cite{FS09}.}

\paragraph{Meaning of $\boldsymbol{\tau_a}$\,.}

There is no $\tau_a$ in the usual T-system, so one may wonder if the system above is equivalent to them. To show the equivalence we consider the following transformation\footnote{We thank Sergey Frolov for the explanation of this subsection.}
\bea
&&T_{a,s}\ \to\  {T_{a,s}\ov F_{a-s+1}F_{a-s+3}\cdots F_{a+s-3}F_{a+s-1}}\,,\qquad s\ge 1\,,
\label{F transf1} \\[1mm]
&&T_{a,s}\ \to\  T_{a,s}\,,\qquad s\le 0\,,
\label{F transf2}
\eea
where $F_a$ are arbitrary functions satisfying
\bea
F_a^+ \, F_a^- = F_{a+1} \, F_{a-1}\,.
\eea
Then it is not difficult to show that under this transformation the Y-functions with $s\neq 0$ are invariant. We can also do a similar transformation with the change $s \to -s$, so there are two independent set of functions $F_a^L$ and $F_a^R$ which can be used for this purpose. $Y_{a,0}$ transforms as 
\bea
Y_{a,0}\ \to\ {\tau_a\ov F_a^L \ssp F_a^R} \, \frac{T_{a,1} \, T_{a,-1}}{T_{a+1,0} \, T_{a-1,0}}\,.
\label{F transf4}
\eea
Then, eqs.\eqref{T system 1} are invariant, and \eqref{T system 0} transforms as
\begin{equation}
T_{a,0}^+ \, T_{a,0}^- = T_{a+1,0} \, T_{a-1,0} + {\tau_a \ov F_a^L \ssp F_a^R} \, T_{a,1} \, T_{a,-1} \,.
\label{T system 0b}
\end{equation}
Thus, if we choose $F_a^L \ssp F_a^R = \tau_a$ we get the T-system in the usual form. This consideration shows that introducing $\tau_a$ is just a matter of convenience.

\section{Asymptotic solutions}\label{app:asymp sol}

\subsection{Asymptotic transfer matrix}

We will discuss the $\su(2|2)$ transfer matrix for the totally symmetric representations on the level-one vacuum, namely when there are no auxiliary Bethe roots among the physical excitations.
Via analytic continuation of the rapidity into the mirror region, this transfer matrix generates the solution of the excite-state TBA equations for the $\sl(2)$ sector \cite{AFSu09} in the asymptotic limit.

As discussed in \cite{GKV09a,AdLST09}, such transfer matrices, here denoted by $T_{a,1}$\,, are given by
\begin{equation}
T_{a,1} = 1 + \frac{\cR^{[-a]}_{(-)} \, \cB^{[-a]}_{(+)}}{\cR^{[+a]}_{(-)} \, \cB^{[+a]}_{(+)}} - 2 \, \frac{\cR^{[+a]}_{(+)}}{\cR^{[+a]}_{(-)}}
+ \sum_{k=1}^{a-1} \( - 2 + \frac{\cR^{[a-2k]}_{(-)}}{\cR^{[a-2k]}_{(+)}} + \frac{\cB^{[a-2k]}_{(-)}}{\cB^{[a-2k]}_{(+)}} \)
\frac{\cR^{[+a]}_{(+)}}{\cR^{[+a]}_{(-)}} \, \frac{\cQ^{[a-1-2k]}}{\cQ^{[a-1]}} \,.
\label{general-tr a}
\end{equation}
This formula consists of $4a$ terms, and only four terms lie outside the sum. It is possible to  include all terms under summation, as
\begin{equation}
T_{a,1} = \sum_{k=0}^{a} \lambda_k^{(B1),a}
+ \sum_{k=1}^{a-1} \lambda_k^{(B2),a}
- \sum_{k=0}^{a-1} \( \lambda_k^{(F1),a} + \lambda_k^{(F2),a} \),
\label{general-tr a2}
\end{equation}
where
\begin{alignat}{5}
\lambda_k^{(B1),a} &= \frac{\cR^{[a-2k]}_{(-)}}{\cR^{[a-2k]}_{(+)}} \, \frac{\cR^{[+a]}_{(+)}}{\cR^{[+a]}_{(-)}} \, \frac{\cQ^{[a-1-2k]}}{\cQ^{[a-1]}} \,, \qquad
\lambda_k^{(B2),a} = \frac{\cB^{[a-2k]}_{(-)}}{\cB^{[a-2k]}_{(+)}} \,
\frac{\cR^{[+a]}_{(+)}}{\cR^{[+a]}_{(-)}} \, \frac{\cQ^{[a-1-2k]}}{\cQ^{[a-1]}} \,,
\notag \\[2mm]
\lambda_k^{(F1),a} &= \lambda_k^{(F2),a} = \frac{\cR^{[+a]}_{(+)}}{\cR^{[+a]}_{(-)}} \, \frac{\cQ^{[a-1-2k]}}{\cQ^{[a-1]}} \,.
\label{def:lambda comp}
\end{alignat}

The auxiliary variables in the vertical direction are defined by taking four terms out of $T_{a,1}$ as
\begin{align}
A_{a,1} &= \lambda_0^{(B1),a} + \lambda_1^{(B2),a}
- \lambda_0^{(F1),a} - \lambda_0^{(F2),a}
= \frac{\cR_{(+)}^{[+a]}}{\cR_{(-)}^{[+a]}} \(
\frac{\cR_{(-)}^{[+a]}}{\cR_{(+)}^{[+a]}}
+ \frac{\cR_{(+)}^{[a-2]}}{\cR_{(-)}^{[a-2]}} - 2 \),
\notag \\[2mm]
\olA_{a,1} &= \lambda_a^{(B1),a} + \lambda_{a-1}^{(B2),a}
- \lambda_{a-1}^{(F1),a} - \lambda_{a-1}^{(F2),a}
= \frac{\cQ^{[-a+1]}}{\cQ^{[+a-1]}} \frac{\cR_{(+)}^{[+a]}}{\cR_{(-)}^{[+a]}} \(
\frac{\cB_{(+)}^{[-a]}}{\cB_{(-)}^{[-a]}}
+ \frac{\cB_{(-)}^{[-a+2]}}{\cB_{(+)}^{[-a+2]}} - 2 \) .
\label{def:asymp A's}
\end{align}
It is easy to check that the covariant recursions \eqref{TA rec XXX} are satisfied for $a \ge 2$, under the condition $\fX_{a,1} = 0$.\footnote{Furthermore, for each sum of \eqref{general-tr a2} one can solve the recursion, {\it e.g.} $T_{a,1}^{(B1) \,+} - A_{a,1}^{(B1) \,+} = \frac{\olA_{a,1}^{(B1) \,+} }{\olA_{a-1,1}^{(B1)} } \, T_{a-1,1}^{(B1)}$\,.}
As a corollary, we find
\begin{equation}
T_{a,0} \, T_{a,2} = A_{a,1}^+ \, \olA_{a,1}^- \,, \qquad (a \ge 2),
\end{equation}
in consistency with \eqref{Tjpm ev}.

\bigskip
The transfer matrices in the horizontal direction $T_{1,s}$ are generated by Bazhanov-Reshetikhin formula \cite{BR89} (see also \cite{Beisert06b,GKV09a,AdLST09}):
\begin{alignat}{3}
T_{a,s} &\equiv \det \begin{pmatrix}
T_{a,1}^{[-s+1]} & T_{a-1,1}^{[-s+2]} & \cdots & T_{a+2-s,1}^{[-1]} & T_{a+1-s,1}  \\[1mm]
T_{a+1,1}^{[-s+2]} & T_{a,1}^{[-s+3]} & \cdots & T_{a+3-s,1} & T_{a+2-s,1}^{[+1]}  \\[1mm]
\vdots & \vdots & & \vdots & \vdots \\[1mm]
T_{a-2+s,1}^{[-1]} & T_{a-3+s,1} & \cdots & T_{a,1}^{[s-3]} & T_{a-1,1}^{[s-2]} \\[1mm]
T_{a-1+s,1} & T_{a-2+s,1}^{[+1]} & \cdots & T_{a+1,1}^{[s-2]} & T_{a,1}^{[s-1]}
\end{pmatrix},
\notag \\[1mm]
T_{a,0} &= 1, \qquad T_{0,s} = 1, \qquad T_{a<0,s} = 0.
\label{Tas-antisymm}
\end{alignat}
Note that the boundary conditions $T_{a,0}=T_{0,s}=1$ are important in applying this formula.
From explicit computation we obtain the following results:
\begin{align}
T_{1,s} &\equiv \Theta_{1,s} \, \TH_{1,s} \,, \qquad
\Theta_{1,s} = (-1)^s \( \prod_{k=1}^{s-1} \frac{\cR_{(+)}^{[-s+2k]} }{\cR_{(-)}^{[-s+2k]} } \),
\label{T1s formula} \\[1mm]
\TH_{1,s} &\equiv (s+1) \, \frac{\cR_{(+)}^{[+s]} }{\cR_{(-)}^{[+s]} }
- s
- s \, \frac{\cR_{(+)}^{[+s]} }{\cR_{(-)}^{[+s]} } \, \frac{\cB_{(+)}^{[-s]} }{\cB_{(-)}^{[-s]} }
+ (s-1) \, \frac{\cB_{(+)}^{[-s]} }{\cB_{(-)}^{[-s]} } \,,
\label{def:TH1s}
\end{align}
for $s \ge 1$. We have $T_{1,0} = 1$ at $s=0$. The factor $\Theta_{1,s}$ satisfies
\begin{equation}
\Theta_{1,s}^- \, \Theta_{1,s}^+ = \Theta_{1,s-1} \, \Theta_{1,s+1} \,,
\qquad (s \ge 2),
\end{equation}
and can be regarded as an artifact of our gauge choice.\footnote{Here $\Theta_{1,s}$ means the gauge transformation from $T_{1,s}$ in \eqref{Tas-antisymm} to $\TH_{1,s}$ in \eqref{def:TH1s}. In the main text, $\Phi_{1,s}$ means a general gauge transformation.}

The gauge choice $T_{1,s} = \TH_{1,s}$ is useful to discuss the horizontal strip of $\alg{su}(2|4|2)$-hook. Note that $\TH_{1,0} \neq 1$ in this gauge. To see this, consider gauge transformation of the T-system at $(1,1)$,
\begin{equation}
\Theta_{1,1}^- \, \Theta_{1,1}^+ \, \TH_{1,1}^- \, \TH_{1,1}^+ =
\Theta_{1,2} \, \Theta_{1,0} \, \TH_{1,2} \, \TH_{1,0} 
+ \Theta_{2,1} \, \Theta_{0,1} \, \TH_{2,1} \, \TH_{0,1} \,,
\end{equation}
where $\Theta_{1,0} \, \TH_{1,0} = 1$, because the original transfer matrix satisfies $T_{1,0}=1$.
Gauge-covariance requires
\begin{equation}
1 = \Theta_{1,1}^- \, \Theta_{1,1}^+ = \Theta_{1,2} \, \Theta_{1,0} = \Theta_{2,1} \, \Theta_{0,1} \,,
\quad \Rightarrow \quad
\Theta_{1,0} = \frac{1}{\Theta_{1,2} } = \frac{\cR_{(-)} }{\cR_{(+)} } \,,
\label{covariance Phi1}
\end{equation}
showing that $\TH_{1,0}=1/\Theta_{1,0} \neq 1$.

\subsection{Asymptotic Wronskian formula}\label{app:asymp Wronskian}

We compare the Wronskian formula \eqref{Wronskian T1s} and the asymptotic transfer matrix $T_{1,s}$ in the $\alg{sl}(2)$ sector to find the asymptotic form of Q-functions. The result should agree with \cite{GKLT10} modulo gauge transformation.

\bigskip
We begin with the relation
\begin{equation}
T_{0,s} \, T_{2,s}
= \frac{\cQ^{[+s-2]}}{\cQ^{[-s+2]}} \, \frac{\cR_{(-)}^{[-s+1]} }{\cR_{(+)}^{[-s+1]} } \, \frac{\cR_{(-)}^{[s-1]} }{\cR_{(+)}^{[s-1]} } \,
(\Theta_{1,s} \, A_{s,1})^+ \, (\Theta_{1,s} \, \olA_{s,1})^-
= L_{1,s}^+ \, \olL_{1,s}^- \,.
\end{equation}
There are many ways to define $L_{1,s}$ and $\olL_{1,s}$\,. We choose the definition such that both $L_{1,s}/\Theta_{1,s}$ and $\olL_{1,s}/\Theta_{1,s}$ are translationally invariant $(X_{1,s} = X_{1,s-1}^+ \,,\ \olX_{1,s} = \olX_{1,s-1}^-)$, as
\begin{alignat}{7}
\frac{L_{1,s}^+ }{\Theta_{1,s}^+ } &= A_{s,1}^+
& &= 1 + \frac{\cR_{(+)}^{[s+1]}}{\cR_{(-)}^{[s+1]}} \, \frac{\cR_{(+)}^{[s-1]}}{\cR_{(-)}^{[s-1]}} - 2 \, \frac{\cR_{(+)}^{[s+1]}}{\cR_{(-)}^{[s+1]}} \,,
\notag \\[2mm]
\frac{\olL_{1,s}^- }{\Theta_{1,s}^- } &= \frac{\cQ^{[+s-2]}}{\cQ^{[-s+2]}} \, \frac{\cR_{(-)}^{[-s+1]} }{\cR_{(+)}^{[-s+1]} } \, \frac{\cR_{(-)}^{[s-1]} }{\cR_{(+)}^{[s-1]} } \, \olA_{s,1}^-
& &= 1 + \frac{\cB_{(+)}^{[-s-1]}}{\cB_{(-)}^{[-s-1]}} \, \frac{\cB_{(+)}^{[-s+1]}}{\cB_{(-)}^{[-s+1]}}
- 2 \, \frac{\cB_{(+)}^{[-s+1]}}{\cB_{(-)}^{[-s+1]}} \,,
\label{asymp L's}
\end{alignat}
where we used the explicit form of $A$'s \eqref{def:asymp A's}.

We define auxiliary variables $(C_{1,s} \,, \olC_{1,s})$ by
\begin{gather}
C_{1,s} = \Theta_{1,s} \( \frac{\cR_{(+)}^{[+s]} }{\cR_{(-)}^{[+s]} } - 1 \),
\qquad
\ol{C}_{1,s} = \Theta_{1,s} \( 1 - \frac{\cB_{(+)}^{[-s]} }{\cB_{(-)}^{[-s]} } \),
\label{def:CC}
\end{gather}
and $(U_{1,s} \,, \olU_{1,s})$ by
\begin{equation}
L_{1,s}^+ = \frac{1}{\Theta_{1,s}^-} \, {\rm det}
\begin{pmatrix}
C_{1,s}^+ & U_{1,s}^+ \\
C_{1,s}^- & U_{1,s}^-
\end{pmatrix},
\qquad
\olL_{1,s}^- = \frac{1}{\Theta_{1,s}^+} \, {\rm det}
\begin{pmatrix}
\olC_{1,s}^+ & \olU_{1,s}^+ \\
\olC_{1,s}^- & \olU_{1,s}^-
\end{pmatrix}.
\label{LL by det}
\end{equation}
The explicit form of $(U_{1,s} \,, \olU_{1,s})$ can be obtained by solving the difference equations \eqref{LL by det}.
The general solutions of these difference equations are
\begin{align}
U_{1,s} &= - \Theta_{1,s} - \pare{ \frac{g}{2 \ssp i} \(v+\frac{is}{g} \) + F_{1,s} } C_{1,s} \,,
\qquad F_{1,s}^- = F_{1,s}^+ \,,
\notag \\[2mm]
\olU_{1,s} &= + \Theta_{1,s} - \pare{ \frac{g}{2 \ssp i} \(v-\frac{is}{g} \) + \olF_{1,s} } \olC_{1,s} \,,
\qquad \olF_{1,s}^+ = \olF_{1,s}^- \,.
\label{UU:asymptotic}
\end{align}
We constrain two periodic functions $F \,, \olF$ by comparing $T_{1,s} = \Theta_{1,s} \, \TH_{1,s}$ in \eqref{def:TH1s} with the Wronskian formula,
\begin{equation}
\TH_{1,s} = \frac{1}{\Theta_{1,s} } \Big( C_{1,s} \, \olU_{1,s} - \olC_{1,s} \, U_{1,s} \Big),
\label{identify T1sQ}
\end{equation}
which gives
\begin{equation}
\Psi_{1,s} \equiv F_{1,s} = \olF_{1,s} \,, \qquad
\Psi_{1,s}^- = \Psi_{1,s}^+ \,.
\label{constraint FF}
\end{equation}
The variable $\Psi$ corresponds to the freedom of superposing two linearly independent solutions $U \to U + \Psi \, C$. The Wronskian formulae \eqref{LL by det}, \eqref{identify T1sQ} do not change as long as $\Psi^+ = \Psi^-$, so we may set $\Psi_{1,s}=0$. Therefore, the asymptotic formula \eqref{UU:asymptotic} can be summarized as
\begin{equation}
U_{1,s} = - \Theta_{1,s} -\frac{g}{2 \ssp i} \(v+\frac{is}{g} \) C_{1,s} \,,
\qquad
\olU_{1,s} = + \Theta_{1,s} - \frac{g}{2 \ssp i} \(v-\frac{is}{g} \) \olC_{1,s} \,.
\label{UU:asymptotic2}
\end{equation}

\bigskip
Let us relate $(C,L,U)$ and the conjugates to the Q-functions in Section \ref{sec:TQ Wronskian}. We can identify\footnote{Such identification of $C$'s is motivated by the expression of \cite{GKLT10}. Note that their definition of $x(v)$ in the mirror region is opposite to ours.}
\begin{gather}
\begin{pmatrix}
C_{1,s} \\[1mm]
\olC_{1,s} \\[1mm]
U_{1,s} \\[1mm]
\olU_{1,s}
\end{pmatrix} \simeq
\begin{pmatrix}
{\sf Q}_{1}^{[+s]} \\[1mm]
{\sf Q}_{\overline{2}}^{[-s]} \\[1mm]
{\sf Q}_{2}^{[+s]} \\[1mm]
{\sf Q}_{\overline{1}}^{[-s]}
\end{pmatrix} =
\begin{pmatrix}
Q_{1,s}^{\rm I} \\[1mm]
\olQ_{1,s}^{\rm I} \\[1mm]
Q_{1,s}^{\rm II} \\[1mm]
\olQ_{1,s}^{\rm II}
\end{pmatrix}, \qquad
\begin{pmatrix}
L_{1,s} \\[1mm]
\olL_{1,s}
\end{pmatrix}
\simeq \begin{pmatrix}
{\sf Q}_{12}^{[s-1]} \\[1mm]
{\sf Q}_{\overline{\emptyset}}^{[-s+1]} \, {\sf Q}_{\overline{12}}^{[-s+1]}
\end{pmatrix}.
\label{identify CUL Q}
\end{gather}
where $\simeq$ means that they are equal up to gauge transformation. One can also check that $(C,L,U)$ and the conjugates solve the difference equations \eqref{covariant Q difference}.
The determinant formula \eqref{LL by det} turns out to be identical to \eqref{Q12 det} if ${\sf Q}_{\overline{\emptyset}} = - 1$.

\subsection{Analyticity data}\label{app:analyticity asymp}

We enlist the analyticity data of the asymptotic solution in the $\sl(2)$ sector. Since the two pairs of fundamental Q-functions $(C_{1,s} \,, \olC_{1,s})$ and $(U_{1,s} \,, \olU_{1,s})$ have almost the same analytic structure (like the location of poles and branch cuts), we denote them collectively by $(Q_{1,s}^\alpha \,, \olQ_{1,s}^\alpha)$.

The location of poles is:
\begin{alignat}{7}
\TH_{1,s} (v) &= \infty \quad {\rm at} \ \ v = u_j - \frac{i (s+1)}{g} \,,
\notag \\[2mm]
L_{1,s} (v) &= \infty \quad {\rm at} \ \ v = u_j - \frac{i (s-1)}{g} \,,
\notag \\[2mm]
Q_{1,s}^\alpha (v) &= \infty \quad {\rm at} \ \ v = u_j - \frac{i (s+1)}{g} \,.
\end{alignat}
The conjugate variables $\olL_{1,s} \,, \olQ_{1,s}^\alpha$ are not singular on the top sheet of the complex plane for $v$.
The location of branch cuts is:
\begin{alignat}{7}
\TH_{1,s} (v+i0) &\neq \TH_{1,s} (v-i0) &\quad &{\rm at} \ \ {\rm Im} \, v = \pm \frac{s}{g} \,,
\notag \\[2mm]
L_{1,s} (v+i0) &\neq L_{1,s} (v-i0) &\quad &{\rm at} \ \ {\rm Im} \, v = - \frac{s-2}{g} \,, \ - \frac{s}{g} \,,
\notag \\[2mm]
\olL_{1,s} (v+i0) &\neq \olL_{1,s} (v-i0) &\quad &{\rm at} \ \ {\rm Im} \, v = + \frac{s-2}{g} \,, \ + \frac{s}{g} \,,
\notag \\[2mm]
Q_{1,s}^\alpha (v+i0) &\neq Q_{1,s}^\alpha (v-i0) &\quad &{\rm at} \ \ {\rm Im} \, v = - \frac{s}{g} \,,
\notag \\[2mm]
\olQ_{1,s}^\alpha (v+i0) &\neq \olQ_{1,s}^\alpha (v-i0) &\quad &{\rm at} \ \ {\rm Im} \, v = + \frac{s}{g} \,,
\end{alignat}
In the limit ${\rm Re} \, v \to \pm \infty$, these functions approach a constant. In short, if $s \ge 3$, the quantities $(L \,, Q)$ are analytic in the upper half plane, whereas $(\olL \,, \olQ)$ are analytic in the lower half plane. One has to be careful about $(L_{1,2} \,, Q_{1,1}^-)$ and $(\olL_{1,2} \,, \olQ_{1,1}^+)$ on the real axis.

\smallskip
As for the variables $(\fb_s^\alpha \,, \fbb_s^\alpha)$, we find the branch cuts at:
\begin{alignat}{7}
\fb_s^\alpha (v+i0) &\neq \fb_s^\alpha (v-i0) &\quad &{\rm at} \ \ {\rm Im} \, v = \pm \frac{s-1}{g} \,, \ - \frac{s+1}{g} \,,
\notag \\[2mm]
\fbb_s^\alpha (v+i0) &\neq \fbb_s^\alpha (v-i0) &\quad &{\rm at} \ \ {\rm Im} \, v = \pm \frac{s-1}{g} \,, \ + \frac{s+1}{g} \,.
\end{alignat}
They do not have poles.

We find extra zeroes for $U_{1,s} \,, \olU_{1,s}$
\begin{equation}
U_{1,s} (v) = 0 \quad {\rm at} \ \ v = - \frac{i (r+s)}{g} \ \ \ (r > 0),
\qquad
\olU_{1,s} (v) = 0 \quad {\rm at} \ \ v = \frac{i (\olr+s)}{g} \ \ \ (\olr > 0),
\end{equation}
and similarly for $(\fb_s^{\rm II} \,, \fbb_s^{\rm II})$.
Both $r, \olr$ are far from the origin already at weak coupling, and they run away from the real axis as $g$ increases.
For other functions we do not find extra zeroes, at least in the weak coupling.

\end{document}